# Characterization of Piezoelectric Materials for Transducers

## Stewart Sherrit

*Jet Propulsion Laboratory, California Institute of Technology, Pasadena, California, USA, ssherrit@jpl.nasa.gov*

## Binu K. Mukherjee

*Physics Department, Royal Military College of Canada, Kingston, ON, Canada, K7K 7B4. mukherjee@rmc.ca*

## 1.1 Introduction

Implementing piezoelectric materials as actuators, vibrators, resonators, and transducers requires the availability of a properties database and scaling laws to allow the actuator or transducer designer to determine the response under operational conditions. A metric for the comparison of the material properties with other piezoelectric materials and devices is complicated by the fact that these materials display a variety of nonlinearities and other dependencies that obscure the direct comparisons needed to support users in implementing these materials. In selecting characterization techniques it is instructive to look at how the material will be used and what conditions it will be subjected to. However in order to define metrics for the higher order effects we need to first quantify the ideal linear behavior of these materials to generate a baseline for the comparisons.

A phenomenological model derived from thermodynamic potentials mathematically describes the property of piezoelectricity. The derivations are not unique and the set of equations describing the piezoelectric effect depends on the choice of potential and the independent variables used. An excellent discussion of these derivations is found in (Mason 1958). In the case of a sample under isothermal and adiabatic conditions and ignoring higher order effects, the elastic Gibbs function may be described by

$$G_1 = -\tfrac{1}{2}\left(s^D_{ijkl}T_{ij}T_{kl} + 2g_{nij}D_nT_{ij}\right) + \tfrac{1}{2}\beta^T_{mn}D_mD_n\,, \tag{0.1}$$

where $g$ is the piezoelectric voltage coefficient, $s$ is the elastic compliance, and $\beta$ is the inverse permittivity. The independent variables in this equation are the stress $T$ and the electric displacement $D$. The superscripts of the constants designate the independent variable that is held constant when defining the constant, and the subscripts define tensors that take into account the anisotropic nature of the material. The linear equations of piezoelectricity for this potential are determined from the derivative of $G_1$ and are

$$\begin{aligned} S_{ij} &= -\frac{\partial G_1}{\partial T_{ij}} = s^D_{ijkl}T_{kl} + g_{nij}D_n \\ E_m &= \frac{\partial G_1}{\partial D_m} = \beta^T_{mn}D_n - g_{nij}T_{ij}\,, \end{aligned} \tag{0.2}$$

where $S$ is the strain and $E$ is the electric field. The above equations are usually simplified to a reduced form by noting that there is a redundancy in the strain and stress variables (Nye 2003) or (Cady 1964) for a discussion detailing tensor properties of materials). The elements of the tensor are contracted to a $6 \times 6$ matrix with 1, 2, 3 designating the normal stress and strain and 4, 5, and 6 designating the shear stress and strain elements. Other representations of the linear equations of piezoelectricity derived from the other possible thermodynamic potentials are shown below (Ikeda 1990, Mason 1958). These sets of equations (0.3)-(0.6) includes equation (0.2) in contracted notation.

$$S_p = s^D_{pq} T_q + g_{pm} D_m$$
$$E_m = \beta^T_{mn} D_n - g_{pm} T_p \tag{0.3}$$

$$S_p = s^E_{pq} T_q + d_{pm} E_m$$
$$D_m = \varepsilon^T_{mn} E_n + d_{pm} T_p \tag{0.4}$$

$$T_p = c^E_{pq} S_q - e_{pm} E_m$$
$$D_m = \varepsilon^S_{mn} E_n + e_{pm} S_p \tag{0.5}$$

$$T_p = c^D_{pq} S_q - h_{pm} D_m$$
$$E_m = \beta^S_{mn} D_n - h_{pm} S_p, \tag{0.6}$$

where *d*, *e*, *g*, and *h* are piezoelectric constants, *s* and *c* are the elastic compliance and stiffness, and ε and β are the permittivity and the inverse permittivity. The relationship described by each of these equations can be represented in matrix form as shown below for equation (0.4):

$$\begin{bmatrix} S_1 \\ S_2 \\ S_3 \\ S_4 \\ S_5 \\ S_6 \\ D_1 \\ D_2 \\ D_3 \end{bmatrix} = \begin{bmatrix} s^E_{11} & s^E_{12} & s^E_{13} & s^E_{14} & s^E_{15} & s^E_{16} & d_{11} & d_{12} & d_{13} \\ s^E_{12} & s^E_{22} & s^E_{23} & s^E_{24} & s^E_{25} & s^E_{26} & d_{21} & d_{22} & d_{23} \\ s^E_{13} & s^E_{23} & s^E_{33} & s^E_{34} & s^E_{35} & s^E_{36} & d_{31} & d_{32} & d_{33} \\ s^E_{14} & s^E_{24} & s^E_{34} & s^E_{44} & s^E_{45} & s^E_{46} & d_{41} & d_{42} & d_{43} \\ s^E_{15} & s^E_{25} & s^E_{35} & s^E_{45} & s^E_{55} & s^E_{56} & d_{51} & d_{52} & d_{53} \\ s^E_{16} & s^E_{26} & s^E_{36} & s^E_{46} & s^E_{56} & s^E_{66} & d_{61} & d_{62} & d_{63} \\ d_{11} & d_{12} & d_{13} & d_{14} & d_{15} & d_{16} & \varepsilon^T_{11} & \varepsilon^T_{12} & \varepsilon^T_{13} \\ d_{21} & d_{22} & d_{23} & d_{24} & d_{25} & d_{26} & \varepsilon^T_{12} & \varepsilon^T_{22} & \varepsilon^T_{23} \\ d_{31} & d_{32} & d_{33} & d_{34} & d_{35} & d_{36} & \varepsilon^T_{13} & \varepsilon^T_{23} & \varepsilon^T_{33} \end{bmatrix} \begin{bmatrix} T_1 \\ T_2 \\ T_3 \\ T_4 \\ T_5 \\ T_6 \\ E_1 \\ E_2 \\ E_3 \end{bmatrix} \tag{0.7}$$

The matrix shown in (0.7) is a generalized representation of the equations shown in equation (0.4). Many elements of the matrix in Eq.(0.7) are zero or not independent due to the symmetry of the crystal which reduces the number of independent constants considerably. For example, the poled ferroelectric ceramic PZT has a $C_\infty = C_{6v}$ crystal class. The symmetric reduced matrix in Eq. (0.7) has 2 independent free dielectric permittivities ($\varepsilon^T_{33}$, $\varepsilon^T_{11} = \varepsilon^T_{22}$), 3 independent piezoelectric constants and ($d_{33}$, $d_{31}$, $d_{15}$) and 5 independent elastic constants under short circuit boundary conditions ($s^E_{11} = s^E_{22}$, $s^E_{33}$, $s^E_{44} = s^E_{55}$, $s^E_{12} = s^E_{21}$, $s^E_{13} = s^E_{23}$, and $s^E_{66} = 2\left(s^E_{11} - s^E_{12}\right)$). The reduced matrix expresses the relationship between the material constants and the variables *S*, *T*, *E*, and *D*. Ideally, under small fields and stresses and for materials with low losses within a limited frequency range these constants contain all the information that is required to predict the behavior of the material under the application of stress, strain, electric field, or the application of charge to the sample surface. In practice, however, most materials display dispersion, nonlinearity, and have measurable losses. For the material constants of the matrix representing Eq.(0.7), a more accurate representation for these constants would be to describe them as coefficients with the functional relationship of the form.

$$s^E_{kl} = s^{E'}_{kl}\left(\omega, E_i, T_{ij}, T, t\right) + i s^{E''}_{kl}\left(\omega, E_i, T_{ij}, T, t\right) \tag{0.8}$$

$$d_{ij} = d'_{ij}\left(\omega, E_i, T_{ij}, T, t\right) + i d''_{ij}\left(\omega, E_i, T_{ij}, T, t\right) \tag{0.9}$$

$$\varepsilon^{\mathrm{T}}_{ij} = \varepsilon^{T'}_{ij}\left(\omega, E_i, T_{ij}, T, t\right) + i \varepsilon^{T''}_{ij}\left(\omega, E_i, T_{ij}, T, t\right) \tag{0.10}$$

where the coefficients are written in terms of the real and imaginary (loss) components as a function of the frequency, electric field, stress, temperature, and time. This is a generalized representation, which accounts for the field $E_i$, stress $T_{ij}$ and frequency dependence ω. Equations (0.8)-(0.10) include the dependence on temperature *T* and time *t*. Many practical piezoelectric materials used for transduction are also ferroelectric in nature and have an associated Curie point that marks a phase change (e.g., ferroelectric to paraelectric). In these materials the elastic, piezoelectric, and dielectric properties de-pend on the proximity of the measurement temperature to the Curie temperature. These materials are poled by the application of a field greater than the coercive field $E_C$ (field at which the dipole orientation begins to switch noticeably). The poling requires a finite time, and after the application of the poling field a relaxation occurs which is described by an aging curve, which is generally logarithmic in time *t*. In some materials, a uniaxial stress is applied to aid in the reorientation of the dipoles. Operation of the sample at high field and high stress may therefore accelerate the relaxation and the sample may be partially depoled. These aging/poling/depoling processes require a measurable time and depend on the present and prior conditions to which the material is/was subjected. This time and temperature dependence is closely associated with the ferroelectric nature of the material.

Berlincourt (Berlincourt et al. 1964) noted that a piezoelectric material that can be poled could be considered to be mathematically identical to a biased electrostrictive material (strain is quadratic with field) with the internal field of the poled piezoelectric supplying the bias field. As can be seen from these equations, the material constants of piezoelectric materials may have a variety of dependencies. These may introduce significant errors in device design and operation if it is assumed that the materials are lossless, linear, and frequency independent. Having noted the various dependencies the material coefficients can posses let us turn our attention to the most prevalent deviation from linear theory, loss and dispersion.

## 1.2 Loss, Phase Shift, Attenuation – Complex notation

A variety of authors have used complex material coefficients to account for phase shifts in the AC response of piezoelectric materials (Berlincourt et al. 1964), (Holland 1967), (Smits 1976, 1985), ( Sherrit et al. 1991,1992), (Alemany et al. 1994). Complex coefficients have been found to model impedance spectra quite accurately and subtle effects in measured data can be accounted for by the use of complex constants. It should be noted that we are describing the behavior of the material and not that of a transducer since the addition of a backing layer or load impedance may obscure the phase shifts associated with the elastic, dielectric, and piezoelectric constants of the material. As we will show, these additional layers require the use of transducer models or Finite Element Models (FEM) to determine the full response. In order to understand what a complex representation implies let us look at a general linear system subjected to an AC continuous input $x = x_0\exp(i\omega t)$. The linear relationship of the response y to the input x is simply.

$$y = b\, x_0 \exp(i\omega t) \qquad (0.11)$$

If the constant b relating the input and output is strictly real then y and x are said to be in phase. If on the other hand b is complex, $b = b_r + ib_i$, then y and x are out of phase by a phase angle φ:

$$y = |b|\, x_0 \exp(i\omega t + \varphi)\ , \qquad (0.12)$$

where

$$\varphi = \operatorname{atan}(b_i/b_r) \qquad |b| = (b_r^2 + b_i^2)^{1/2} \qquad (0.13)$$

As well, a distinction should be made between the *macroscopic* behaviors of these materials that are described by the complex components and the *microscopic* behavior that is the source of these complex components and their associated phase shift. In linear piezoelectric theory, the stress T or the electric displacement D of a piezoelectric material subjected to a field E or a strain S excitation is described macroscopically by the coupled linear piezoelectric equations determined from the electric Gibbs' function (Eq. (0.5)) with coefficients $c_{pq}^E$, $e_{pm}$, and $\varepsilon_{mn}^S$, which are respectively the elastic stiffness at constant field, the piezoelectric charge coefficient, and the clamped permittivity. The phase shifts associated with the material coefficients of a piezoelectric material for a single frequency excitation are summarized in each of the subsections below.

### 1.2.1 Elastic Properties

The complex stiffness (anisotropic material) or the Young's Modulus Y (isotropic material) in the strain-stress relationship $T = cS$ describes the propagation and dissipation of ultrasonic waves in a material (McSkimin 1964). The wave equation with a complex stiffness $\mathbf{c} = c_r + ic_i$ is

$$\mathbf{c}\frac{\partial^2 u}{\partial x^2} = \rho\frac{\partial^2 u}{\partial t^2} \tag{0.14}$$

where $u$ is the displacement and ρ is the density. The solution to the wave equation is of the form

$$u(x,t) = u_0 \exp(i(\omega\, t - \Gamma x)) \tag{0.15}$$

and Γ is the propagation constant (complex) and is related to the complex stiffness **c** by

$$\Gamma = \sqrt{\frac{\rho}{\mathbf{c}}}\omega = \frac{\omega}{v_r + iv_i} = \frac{v_r - iv_i}{v_r^2 + v_i^2}\omega \tag{0.16}$$

The constant $\sqrt{\mathbf{c}/\rho} = \mathbf{v} = v_r + iv_i$ is the complex wave velocity and describes both propagation and attenuation in the material. The solution shown in equation (0.15) may be written as

$$u(x,t) = u_0 \exp(i(\omega\, t - \frac{v_r \omega x}{v_r^2 + v_i^2}) \exp(\frac{-v_i \omega x}{v_r^2 + v_i^2}), \tag{0.17}$$

which may be re-written as

$$u(x,t) = u_0 \exp(i\omega\, t - i\beta\, x) \exp(-\alpha\, x) \ , \tag{0.18}$$

where α is the attenuation constant and β is the angular frequency divided by the wave speed. The mechanical Q of the material, defined as the energy stored divided by the energy dissipated over one period, is

$$Q = \frac{c_r}{c_i} \cong \frac{\beta}{2\alpha} \cong \frac{v_r}{2v_i} \ . \tag{0.19}$$

It is interesting to note that using this definition of loss, the $Q$ of the material is independent of frequency and that the attenuation coefficient is linearly proportional to frequency as is found in many solid materials (Mason 1950).

#### *1.2.2 Dielectric Properties*

The use of a complex permittivity $\boldsymbol{\varepsilon} = \varepsilon_r + i\varepsilon_i$ to model the AC response of dielectric materials is standard practice in the dielectric community. Data is commonly reported in terms of the dielectric constant and the dissipation

$$\boldsymbol{\kappa} = \frac{\varepsilon_r}{\varepsilon_0}\left(1 + i\frac{\varepsilon_i}{\varepsilon_r}\right) = \frac{\varepsilon_r}{\varepsilon_0}\left(1 + i\tan\delta\right) = \kappa_r + i\kappa_i \ , \tag{0.20}$$

where the dissipation tanδ can be shown to be the ratio of the leakage current to the charging current. For a linear dielectric, described by D=εE, with a complex permittivity and a DC conductivity σ, the dissipation in the material under the application of an AC electric field $E = E_0\exp(-i\omega t)$ is given by

$$\tan\delta = \frac{\varepsilon_i + \sigma/\omega}{\varepsilon_r} \tag{0.21}$$

For the majority of piezoelectric materials the conductivity is quite low and the frequencies of interest are in the region of kilohertz and higher. As well tanδ is found to have a modest frequency dependence at frequencies of 1 kHz or higher. For typical piezoelectric materials from the ferroelectric ceramics family, the complex permittivity is largely due to a polarization lag in the material and this is found to be much larger than the contribution of the conduction current due to the DC conductivity.

As has been pointed out (Von Hippel 1967) it is better to model the AC response of a dielectric material with a complex permittivity rather than use an RC circuit. Using an RC circuit presumes the conductivity term is dominant and this imposes an inverse frequency dependence in the dissipation data that is not present in the majority of dielectric materials in the frequency range of interest. This does not mean that a frequency dependence of the dielectric constant is not present. Dispersion does indeed exist but it is not necessarily an inverse frequency dependence.

### *1.2.3 Piezoelectric Properties*

Experimental evidence has been available for some time which confirms the existence of a loss component of the piezoelectric constant for a variety of different materials. In an interesting set of measurements (Wang et al. 1993) measured the phase angles of different piezoelectric materials using both the direct and the converse piezoelectric effect by monitoring the magnitude and phase of the response to a small AC stress (direct effect) and a small AC field (converse effect) and they were able to show that, for the materials they studied, the phase angle was the same (to the accuracy of the measurement) regardless of the excitation. For the direct effect they found that the electric displacement as a function of the AC stress behaved as

$$D_3 = (d_{33r} + id_{33i}) T_0 e^{i\omega t} = |d_{33}| T_0 e^{i(\omega t + \theta_d)} \tag{0.22}$$

whereas the converse effect could be described by

$$S_3 = (d_{33r} + id_{33i}) E_0 e^{i\omega t} = |d_{33}| E_0 e^{i(\omega t + \theta_d)} \tag{0.23}$$

where $\theta_d$ is the phase angle of the electric displacement and strain compared to the driving stress or electric field signals and $|d_{33}|$ is the magnitude of the piezoelectric constant.

### *1.2.4 Electromechanical Properties*

The electromechanical coupling constants of the various piezoelectric resonators have been determined for the case of loss-less materials (Berlincourt et al. 1964). In the case of the thickness resonator the electromechanical coupling constant is defined as

$$k_t^2 = \frac{e_{33}^2}{c_{33}^D \varepsilon_{33}^S} \tag{0.24}$$

And in the low frequency limit is a measure of the conversion efficiency between the energy density of the applied mechanical excitation and the available energy density of the electrical signal or vice versa. In the case of loss-less materials the two energy densities are in phase. In the case of materials with losses, as defined above, a phase shift in the two energy densities occurs and the energy conversion can be shown to lag the excitation by treating the constants as complex and rewriting equation (0.24) in polar notation. A more detailed discussion of this can be found in a recent article (Lamberti et al. 2005). It should be noted that in resonance the energy densities have a spatial dependence and an integral is required to calculate the total energy densities. The total phase angle between the energy density of the excitation signal and the available energy density is

$$\theta = 2\theta_e - \theta_{c^D} - \theta_{\varepsilon^S}, \tag{0.25}$$

where, in general, the phase angle $\theta_x$ of a constant $\boldsymbol{x}=x_r+ ix_i$ is defined by

$$\theta_x = \arctan\left(\frac{x_i}{x_r}\right) \tag{0.26}$$

A complex coupling implies that the mechanical and electrical energy densities are not in phase and that the maximum of the product of the D-E does not coincide in time with the maximum in the S-T curve and vice versa. Care should be taken when applying energy based definitions of the coupling in lossy materials as the linear equations of piezoelectricity are derive for isothermal and adiabatic conditions. The conversion path and the thermal boundary conditions likely become important as the loss increases since other properties such as heat capacity, thermal expansion, pyroelectricity, piezocaloric and electrocaloric properties may have a larger effect (Nye 2003).

## *1.3 Resonance Equations*

The coupling of the electric field to the mechanical stress that is found in the linear constitutive equations means that a stress wave can be generated in a material at a given frequency ω by driving the material electrically at ω. This driving force in a sample of fixed geometry can be used to create an electrically driven mechanical resonance in the material if the boundary conditions permit. A variety of resonance geometries can be produced which can be reduced to one dimension. Figure 1 shows common resonance geometries that can be used to characterize piezoelectric materials along with the poling direction and recommended aspect ratios. The aspect ratio is chosen to ensure that the modes of resonance are well separated and that the desired motion is primarily in the specified direction.

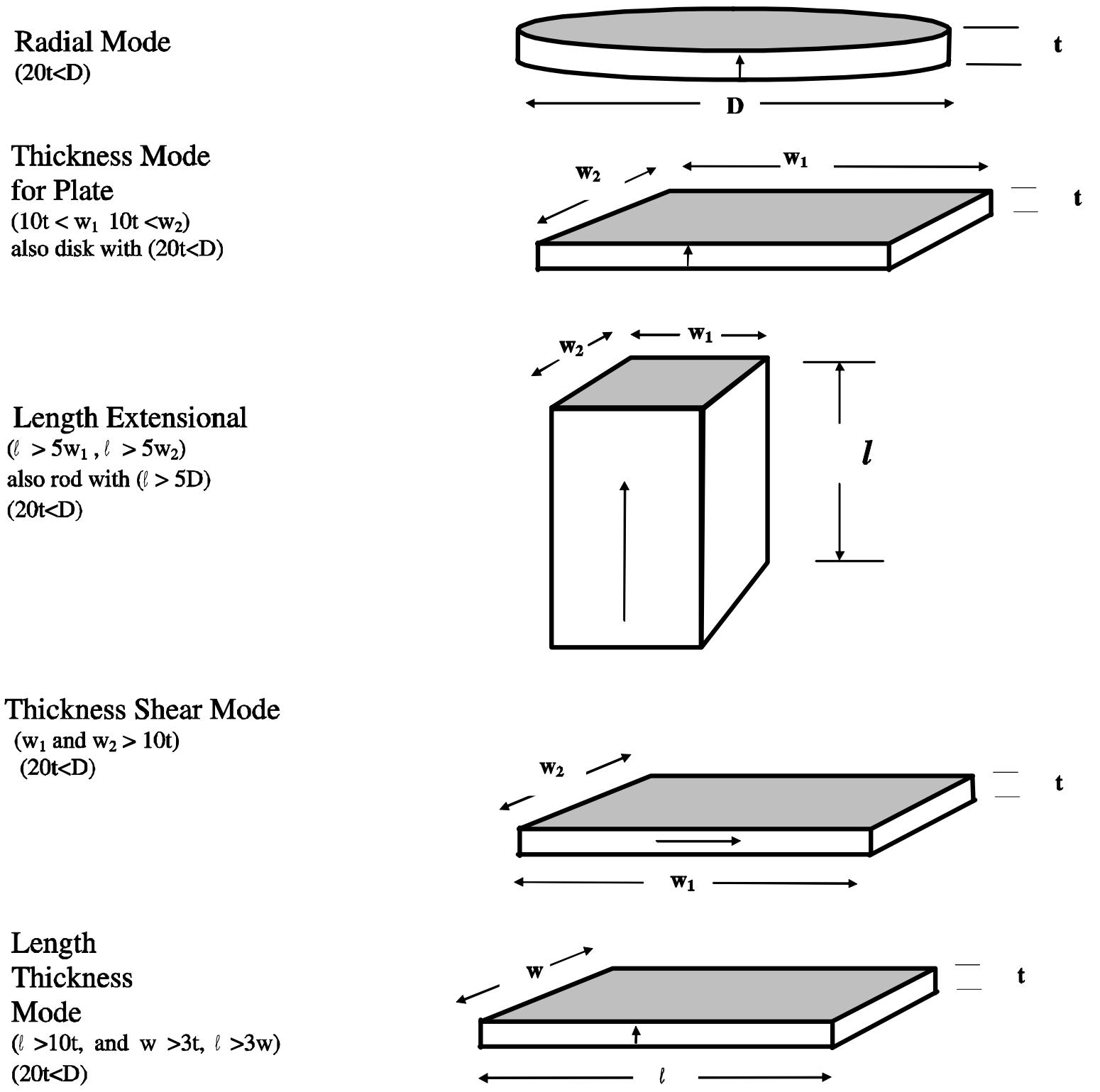

**Fig. 1.** The geometry and poling direction for the five most common modes of piezoelectric resonators used for materials characterization. Note: The radial mode also has a thickness resonance and the length extensional resonator may also be of the form of a long rod.

Consider the case of a plate of poled piezoelectric material with electrodes at the major surfaces of area $A$ and the poling direction perpendicular to the electrode planes which are separated by a distance $l$. The electrode surfaces are traction free and the normal stress in the poling direction is $\mathbf{T}_3 = 0$ at $x_3 = 0$ and $x_3 = l$. In order to investigate the effects of the complex elastic, dielectric and piezoelectric constants on the impedance spectra let us calculate the thickness mode impedance equations using complex coefficients. This derivation is based on that by (Berlincourt et al. 1964) but it is extended to include complex coefficients for the material constants. The constitutive equations governing the thickness resonance are shown in equation(0.5). In an impedance measurement a sinusoidal electric voltage is applied across the sample and the current, including its phase, is measured. The complex impedance is determined by dividing the voltage by the phasor of the current. The voltage and the current have to be calculated under harmonic excitation to determine the impedance equations. The time dependences of the fields T, S, E, and D are taken into account by an exp(-iωt) term and bold typeface is used to denote a complex variable or vector.

$$\mathbf{T}_3(t) \Rightarrow \mathbf{T}_3 e^{-i\omega t} \tag{0.27}$$

$$\mathbf{S}_3(t) \Rightarrow \mathbf{S}_3 e^{-i\omega t} \tag{0.28}$$

$$\mathbf{D}_3(t) \Rightarrow \mathbf{D}_3 e^{-i\omega t} \tag{0.29}$$

$$\mathbf{E}_3(t) \Rightarrow \mathbf{E}_3 e^{-i\omega t} \tag{0.30}$$

The linear piezoelectric equations (0.5) and the wave equation

$$\frac{\partial^2 u_3}{\partial t^2} = \frac{\mathbf{c}_{33}^D}{\rho} \frac{\partial^2 u_3}{\partial x_3^2} \tag{0.31}$$

are coupled since the strain is defined as

$$\mathbf{S}_3 = \frac{\partial u_3}{\partial x_3} \ , \tag{0.32}$$

where $u_3$ is the displacement along the 3 direction ($x_3$). The general solution of the wave equation is of the form.

$$u_3 = \left[ \boldsymbol{A}\sin\left(\frac{\omega x_3}{\mathbf{v}_D}\right) + \boldsymbol{B}\cos\left(\frac{\omega x_3}{\mathbf{v}_D}\right)\right] e^{-i\omega t} \tag{0.33}$$

where

$$\mathbf{v}_D = \sqrt{\frac{\mathbf{c}_{33}^D}{\rho}} \ . \tag{0.34}$$

Taking the derivative with respect to $x_3$, we have

$$\mathbf{S}_3 = \frac{\partial u_3}{\partial x_3} = \frac{\omega}{\mathbf{v}_D}\left[ \boldsymbol{A}\cos\left(\frac{\omega x_3}{\mathbf{v}_D}\right) - \boldsymbol{B}\sin\left(\frac{\omega x_3}{\mathbf{v}_D}\right)\right] e^{-i\omega t} \tag{0.35}$$

Substituting this equation into the first linear equation (0.5) for a linear piezoelectric material and using the first boundary condition **($\mathbf{T}_3$** =0 at $x_3$ =0), we get

$$0 = \left[ \mathbf{c}_{33}^{\mathbf{D}} \left( \frac{\omega}{\mathbf{v}_D}\left[ \boldsymbol{A}\cos(0) - \boldsymbol{B}\sin(0)\right]\right) - \mathbf{h}_{33}\mathbf{D}_3 \right] e^{-i\omega t} \tag{0.36}$$

which gives

$$\boldsymbol{A} = \frac{\mathbf{h}_{33}\mathbf{D}_3\mathbf{v}_D}{\mathbf{c}_{33}^D\omega} \tag{0.37}$$

At ($\mathbf{T}_3$ =0 at $x_3$ =$l$), where $l$ is the sample thickness, the other boundary condition is

$$0 = \left( \mathbf{c}_{33}^{\mathbf{D}} \left( \frac{\omega}{\mathbf{v}_D}\left[ \boldsymbol{A}\cos\left(\frac{\omega l}{\mathbf{v}_D}\right) - \boldsymbol{B}\sin\left(\frac{\omega l}{\mathbf{v}_D}\right)\right]\right) - \mathbf{h}_{33}\mathbf{D}_3 \right) e^{-i\omega t} \tag{0.38}$$

Rearranging, we find

$$\boldsymbol{B} = \frac{\mathbf{h}_{33}\mathbf{D}_3\mathbf{v}_D}{\mathbf{c}_{33}^D\omega} \frac{\cos\left(\frac{\omega l}{\mathbf{v}_D}\right) - 1}{\sin\left(\frac{\omega l}{\mathbf{v}_D}\right)} . \tag{0.39}$$

Now knowing A and B explicitly $\mathbf{S}_3$ equals

$$\mathbf{S}_3 = \left[ \frac{\mathbf{h}_{33}\mathbf{D}_3}{\mathbf{c}_{33}^D} \left( \cos\left(\frac{\omega x_3}{\mathbf{v}_D}\right) - \frac{\cos\left(\frac{\omega l}{\mathbf{v}_D}\right) - 1}{\sin\left(\frac{\omega l}{\mathbf{v}_D}\right)} \sin\left(\frac{\omega x_3}{\mathbf{v}_D}\right)\right)\right] e^{-i\omega t} . \tag{0.40}$$

Using the second of the linear equations (0.5) the voltage on the sample is

$$\mathbf{V} = -\int_0^l \mathbf{E}_3 dx_3 = \int_0^l \left( \mathbf{h}_{33}\mathbf{S}_3 - \boldsymbol{\beta}_{33}^s \mathbf{D}_3 \right) dx_3 \tag{0.41}$$

$$\mathbf{V}=\int_0^l\left(\mathbf{h}_{33}\left[\frac{\mathbf{h_{33}D_3}}{\mathbf{c}_{33}^D}\left(\cos\left(\frac{\omega x_3}{\mathbf{v}_D}\right)-\frac{\cos\left(\frac{\omega l}{\mathbf{v}_D}\right)-1}{\sin\left(\frac{\omega l}{\mathbf{v}_D}\right)}\sin\left(\frac{\omega x_3}{\mathbf{v}_D}\right)\right)\right]-\boldsymbol{\beta}_{33}^s\mathbf{D}_3\right)dx_3 \tag{0.42}$$

$$\mathbf{V}=-\boldsymbol{\beta}_{33}^s\mathbf{D}_3 l-\frac{\mathbf{h}_{33}^2\mathbf{D_3}\mathbf{v}_D}{\mathbf{c}_{33}^D\omega}\left(\frac{2\cos\left(\frac{\omega l}{\mathbf{v}_D}\right)-2}{\sin\left(\frac{\omega l}{\mathbf{v}_D}\right)}\right) \tag{0.43}$$

Using the identity

$$\tan\left(\frac{x}{2}\right)=\frac{1-\cos(x)}{\sin(x)}, \tag{0.44}$$

the voltage simplifies to

$$\mathbf{V}=-\boldsymbol{\beta}_{33}^s\mathbf{D}_3 l+\frac{\mathbf{h}_{33}^2\mathbf{D_3}\mathbf{v}_D}{\mathbf{c}_{33}^D\omega}\left(2\tan\left(\frac{\omega l}{2\mathbf{v}_D}\right)\right). \tag{0.45}$$

The current may be found using the dielectric displacement

$$\mathbf{I}=A\frac{d\mathbf{D}_3}{dt}=-i\omega A\mathbf{D}_3 \tag{0.46}$$

where $A$ is the electrode area. The impedance of the resonator is finally found using **Z =V/I**

$$\mathbf{Z}=\frac{\boldsymbol{\beta}_{33}^s\mathbf{D}_3 l-\frac{\mathbf{h}_{33}^2\mathbf{D_3}\mathbf{v}_D}{\mathbf{c}_{33}^D\omega}\left(2\tan\left(\frac{\omega l}{2\mathbf{v}_D}\right)\right)}{i\omega A\mathbf{D_3}} \tag{0.47}$$

or

$$\mathbf{Z}=\frac{\boldsymbol{\beta}_{33}^s l-\frac{\mathbf{h}_{33}^2\mathbf{v}_D}{\mathbf{c}_{33}^D\omega}\left(2\tan\left(\frac{\omega l}{2\mathbf{v}_D}\right)\right)}{i\omega A} \tag{0.48}$$

Now, using the relationships

$$\mathbf{k}_t^2=\frac{\mathbf{h}_{33}^2}{\boldsymbol{\beta}_{33}^S\mathbf{c}_{33}^D}\qquad 4\boldsymbol{f}_p=\frac{2\mathbf{v}_D}{l}\qquad \boldsymbol{\beta}_{33}^S=\frac{1}{\boldsymbol{\varepsilon}_{33}^S} \tag{0.49}$$

the equation for the thickness mode (IEEE Standards ANSIIEEE 176 -1987) with complex coefficients is

$$\mathbf{Z}=\frac{l}{i\omega A\boldsymbol{\varepsilon}_{33}^S}\left(1-\frac{\mathbf{k}_t^2\tan\left(\frac{\omega}{4\mathbf{f_p}}\right)}{\frac{\omega}{4\mathbf{f_p}}}\right) \tag{0.50}$$

In the above equation $\boldsymbol{\varepsilon}_{33}^S$, $\boldsymbol{f}_p$, $\boldsymbol{k}_t$ are complex and produce an impedance spectra as a function of frequency $Z(\omega)=R(\omega)+iX(\omega)$ or $Y(\omega)=G(\omega)+iB(\omega)$ that is also complex. It is instructive to look at the effect of the real and imaginary components of the material coefficients on the impedance spectra. Figure 2 shows the admittance divided by the angular frequency ω of a thickness resonator and the impedance components multiplied by frequency R(ω)ω, X(ω)ω .

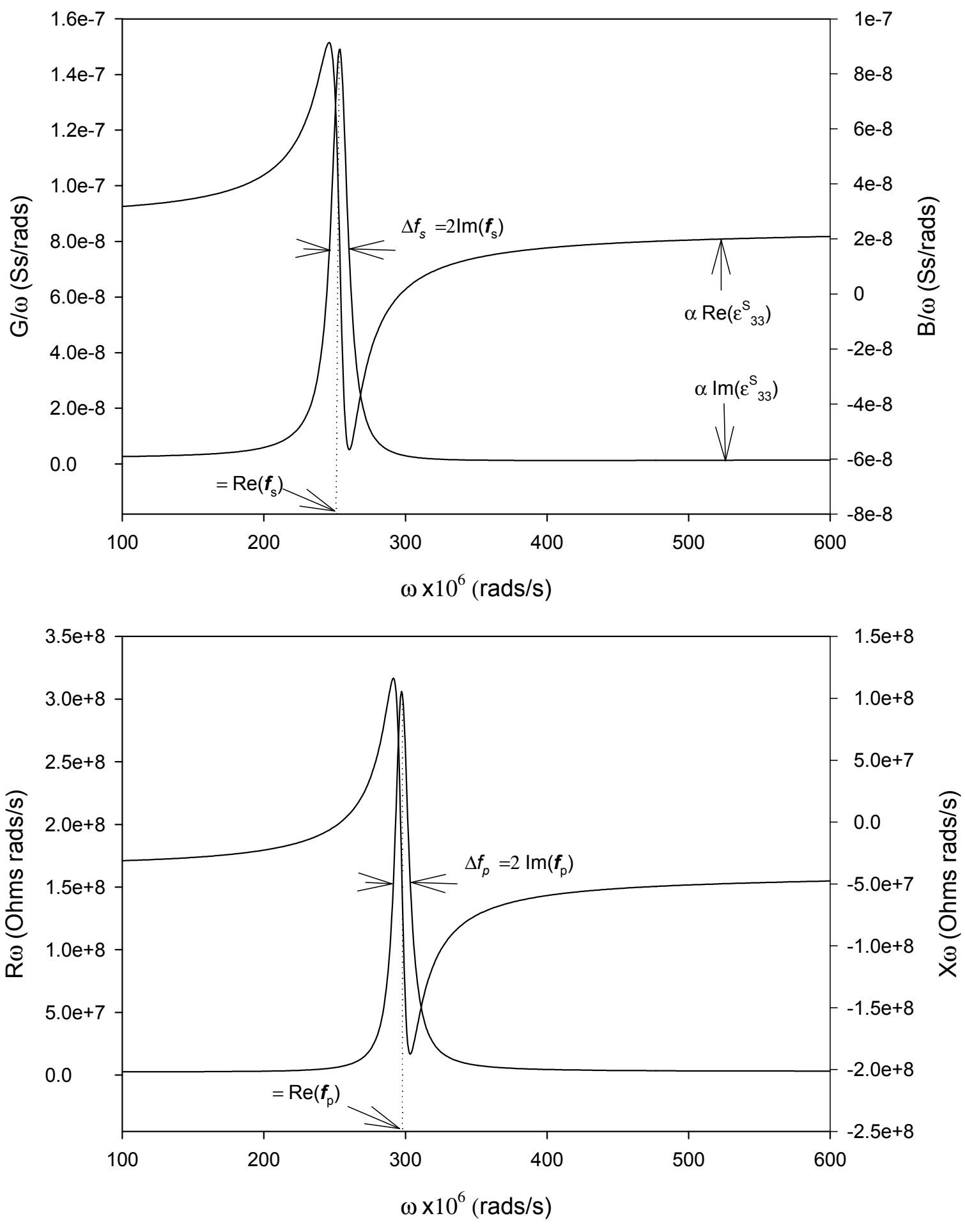

**Fig. 2.** Resonance curves for a thickness mode resonator plotted as G(ω)/ω,B(ω)/ω versus ω and R(ω)ω and X(ω)ω as a function of ω.

The real part of the clamped permittivity $\mathrm{Re}\left(\boldsymbol{\varepsilon}_{33}^{S}\right)$ is proportional to the baseline of the B(ω)/ω spectra above resonance while the imaginary component $\mathrm{Im}\left(\boldsymbol{\varepsilon}_{33}^{S}\right)$ is proportional to the baseline G(ω)/ω spectra above resonance (ideally at $2f_p$) . The complex parallel resonance frequency $\boldsymbol{f}_p$ (maximum in the resistance) is a function of the complex stiffness $\boldsymbol{c}_{33}^{D}$ and the density $\rho$ through the equation $\boldsymbol{c}_{33}^{D} = 4\rho l^2 \boldsymbol{f}_p^2$. The real value of the stiffness is therefore equal to the real part of $\boldsymbol{f}_p$ while the imaginary part of $\boldsymbol{f}_p$ is equal to the half width at half maximum about $f_p$. The final coefficient that can be determined from the spectra is the electromechanical coupling coefficient $\boldsymbol{k}_t$. The real part of the coupling is proportional to the difference of the real part of $\boldsymbol{f}_p$ and the real part of $\boldsymbol{f}_s$ the series resonance frequency (maximum in G). The imaginary part of the coupling is proportional to the ratio of the imaginary part of $\boldsymbol{f}_p$ and the imaginary part of $\boldsymbol{f}_s$ which is a measure of the change in the half width at half maximum at the parallel and series resonance. If the imaginary component of the coupling is zero then by definition $\Delta f_s / f_s = \Delta f_p / f_p$ and the breadth of the resonance in each spectrum is the same.

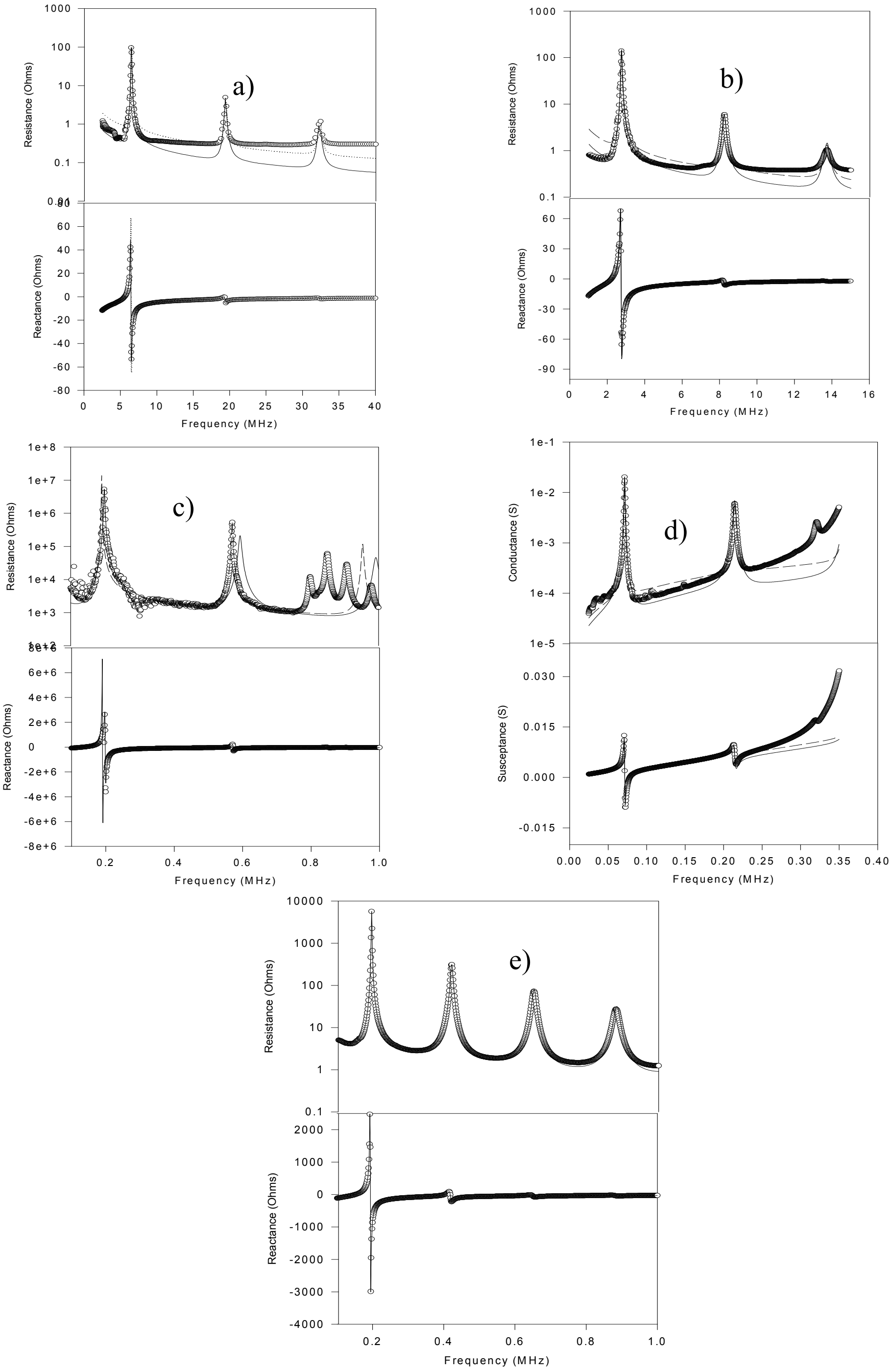

**Fig. 3.** Resonance curves for various modes used to determine the reduced matrix. All data is for the CTS 3203HD material with density ρ= 7800 $Kg/m^3$. a) Thickness extensional TE resonance peaks. b) Thickness shear extensional TSE resonance. c) Length extensional LE resonance. d) Length thickness extensional LTE resonance and the fit. e) Radial extensional RE resonance and fit. The fit to the fundamental resonance peak is shown as a solid line. The fit to the second resonance peak is shown as a broken line.

The thickness mode, although very important in transducer design, is but one of a multitude of one dimensional resonator geometries that can be fabricated and tested. The complex equations for the five common modes used in material characterization are shown in Figure 1 and listed in Table 1. In addition the complex resonance equations for a variety of commercial geometries are listed in Table

2. Figure 3 shows the resonance spectra for the various modes for the sample geometries shown in Figure 1. For a discussion of these and other modes of resonance refer to (Berlincourt et al. 1964). and (Onoe and Jumongi 1967). The resonance frequency is controlled by the acoustic velocity in the material under short circuit conditions and the resonance length of the 1 D geometry. As a general rule of thumb for commercial PZT's, the frequency is about 2 MHz for a millimeter of length in the resonance dimension assuming an acoustic velocity of about 4000 m/s. This corresponds to a frequency constant N = 2000 Hz-m. The strength and the frequency of a particular resonance along with some prior knowledge about the elastic properties can be used to identify a particular resonance if the resonance geometry corresponds to the aspect ratios shown in Figure 1.

### 1.3.1 Resonance Measurement Considerations

In many ways resonance analysis is the simplest method to determine the small signal elastic, dielectric, and piezoelectric properties of a material but there are a variety of points that should be considered if one wants to determine accurate, reliable and repeatable material properties:

a) Know the limits of your instrument. During the sweep of a resonance spectrum the impedance of the sample can change over 6 orders of magnitude between the resonance and anti-resonance frequencies. In order to determine material coefficients accurately, especially the loss terms, make sure the instrument can accurately measure the impedance over this range or alternatively that your analysis does not use data from regions which may not have been measured accurately.

b) If the instruments have open and short circuit corrections for the test-probe or probe compensation these should always be used. This is especially true when custom built test holders are used.

c) The recommended sample aspect ratios must be used to ensure that you have an isolated resonance mode. In the case of extremely anisotropic material properties these recommended aspect ratios may have to be increased.

d) As far as possible the leads to the electrodes of the resonator should be attached at nodal points and, if the contact is not a permanent solder contact, the contact force must be small and the mass of the contact should not be significant. The sample should also be mounted in a symmetric manner with respect to the mode.

e) For thin large lateral dimension resonators such as the thickness, radial and length thickness mode resonators the sheet resistance of the electrodes should be much less than the resistance at resonance.

### 1.3.2 Resonance Analysis

The IEEE Standard on Piezoelectricity (IEEE Standards ANSI/IEEE 176 -1987) uses the impedance equations discussed in the previous section and critical frequencies derived from the equations to determine the real parts of the material constants. The losses associated with the dielectric constant are determined away from resonance, and losses associated with the elastic constants are determined using equivalent circuits recommended by the standard. In the case of high losses, the IEEE Standard recommends the circuit analysis techniques discussed by (Martin 1954) In the current Standard on Piezoelectricity the material constants of the impedance equation are considered to be real. As was shown in the last section the impedance equation for the thickness resonator has the form shown in equation (0.50) where $l$ is the sample thickness, A is the electrode area and $\mathbf{c}_{33}^{D}$, $k_t$ and $\boldsymbol{\varepsilon}_{33}^{S}$ are the elastic stiffness, thickness electromechanical coupling constant and clamped dielectric permittivity. The fundamental anti-resonance in the impedance spectrum (first maximum in the resistance spectra) occurs at $f = f_p$ when the argument of the tangent function in equation (0.50) is equal to $\pi/2$. or when

$$f = \frac{1}{2t}\sqrt{\frac{c_{33}^{D}}{\rho}} = f_p \quad (0.51)$$

**Table 1.** The equations of the common resonance modes used for material characterization of C∞ materials and the associated parameters of the resonance.

| Resonance Mode | Equation | Coefficients relationships |
|---|---|---|
| **Thickness Extensional** (Berlincourt et al. 1964) ( IEEE Std. 176-1987) (Sherrit et al. 1997c) | $\mathbf{Z}(\omega)=\frac{l}{i\omega A\boldsymbol{\varepsilon}_{33}^{S}}\left(1-\frac{\mathbf{k}_t^2\tan\left(\frac{\omega}{4\mathbf{f_p}}\right)}{\frac{\omega}{4\mathbf{f_p}}}\right)$ | $\mathbf{k}_t^2=\frac{\mathbf{h}_{33}^2}{\boldsymbol{\beta}_{33}^S\mathbf{c}_{33}^D}$, $f_p=\frac{\mathbf{v}_D}{2l}=\frac{\sqrt{c_{33}^D/\rho}}{2l}$ <br> $\boldsymbol{\beta}_{33}^S=\frac{1}{\boldsymbol{\varepsilon}_{33}^S}$ |
| **Length Extensional** (Berlincourt et al. 1964) ( IEEE Std. 176-1987) (Sherrit et al. 1997c) | $\mathbf{Z}(\omega)=\frac{l}{i\omega A\boldsymbol{\varepsilon}_{33}^{S}}\left(1-\frac{\mathbf{k}_{33}^2\tan\left(\frac{\omega}{4\mathbf{f_p}}\right)}{\frac{\omega}{4\mathbf{f_p}}}\right)$ | $k_{33}^2=\frac{d_{33}^2}{\varepsilon_{33}^T s_{33}^E}$, $\boldsymbol{\varepsilon}_{33}^S=\left(1-k_{33}^2\right)\boldsymbol{\varepsilon}_{33}^T$ <br> $\mathbf{f}_p=\left[4\left(1-k_{33}^2\right)\rho\mathbf{s}_{33}^{\mathbf{E}}l^2\right]^{-1/2}=\left[4\rho\mathbf{s}_{33}^{D}l^2\right]^{-1/2}$ |
| **Thickness Shear Extensional** (Berlincourt et al. 1964) ( IEEE Std. 176-1987) (Sherrit et al. 1997c) | $\mathbf{Z}(\omega)=\frac{t}{i\omega A\boldsymbol{\varepsilon}_{11}^{S}}\left(1-\frac{\mathbf{k}_{15}^2\tan\left(\frac{\omega}{4\mathbf{f_p}}\right)}{\frac{\omega}{4\mathbf{f_p}}}\right)$ <br> See also Length Shear Resonator (Aurelle 1994), (Pardo 2007) | $k_{15}^2=\frac{d_{15}^2}{\varepsilon_{11}^T s_{55}^E}=\frac{e_{15}^2}{\varepsilon_{11}^s c_{55}^D}$ also $c_{55}^{D,E}=\frac{1}{s_{55}^{D,E}}$ <br> $\mathbf{f}_p=\left[4\left(1-k_{15}^2\right)\rho\mathbf{s}_{55}^{\mathbf{E}}l^2\right]^{-1/2}=\left[4\rho\mathbf{s}_{55}^{D}l^2\right]^{-1/2}$ <br> $\boldsymbol{\varepsilon}_{11}^s=\left(1-k_{15}^2\right)\boldsymbol{\varepsilon}_{11}^T$ |
| **Length Thickness Extensional** (Berlincourt et al. 1964) ( IEEE Std. 176-1987) ( Smits 1976, 1985) | $\mathbf{Y}(\omega)=\left(\frac{i\omega\boldsymbol{\varepsilon}_{33}^T A}{t}\right)\left[1-\mathbf{k}_{13}^2\left[1-\frac{\tan\left(\frac{\omega}{4\mathbf{f}_s}\right)}{\frac{\omega}{4\mathbf{f_s}}}\right]\right]$ | $k_{13}^2=\frac{d_{13}^2}{\varepsilon_{33}^T s_{11}^E}$ <br> $\mathbf{f}_S=\left[4\rho\mathbf{s}_{11}^D l^2/\left(1-k_{13}^2\right)\right]^{-1/2}=\left[4\rho\mathbf{s}_{11}^E l^2\right]^{-1/2}$ |
| **Radial** (Meitzler et al. 1973), ( Sherrit et al. 1991), (Alemany et al. 1995), (Mason 1950) | $\mathbf{Y}(\omega)=\left(\frac{-i\omega\pi a^2\boldsymbol{\varepsilon}_{33}^p}{t}\right)\left[\frac{2(\mathbf{k}^p)^2}{1-\boldsymbol{\sigma}^p-\mathbf{J}(\omega a/(\mathbf{c}_{11}^p/\rho)^{1/2})}-1\right]$ | $\boldsymbol{\varepsilon}_{33}^p\equiv\boldsymbol{\varepsilon}_{33}^T-\frac{2\mathbf{d}_{13}^2}{\mathbf{s}_{11}^E-\mathbf{s}_{12}^E}$ <br> $\mathbf{c}_{11}^p\equiv\frac{\mathbf{s}_{11}^E}{\left(\mathbf{s}_{11}^E\right)^2-\left(\mathbf{s}_{12}^E\right)^2}$ and $\boldsymbol{\sigma}^p\equiv\frac{-\mathbf{s}_{12}^E}{\mathbf{s}_{11}^E}$ <br> $\left(\mathbf{k}^p\right)^2\equiv\frac{\left(\mathbf{e}_{13}^p\right)^2}{\boldsymbol{\varepsilon}_{33}^p\mathbf{c}_{11}^p}$ and $\mathbf{e}_{13}^p\equiv\frac{\mathbf{d}_{13}}{\mathbf{s}_{11}^E+\mathbf{s}_{12}^E}$ |

Rearranging equation (0.51) the elastic stiffness $c_{33}^D$ is given in terms of the parallel resonance frequency by.

$$c_{33}^D=4\rho\,t^2 f_p^2 \tag{0.52}$$

The electromechanical coupling factor can be determined in a similar manner. The series resonance frequency $f_s$ in the admittance spectrum is determined by the maximum in the conductance versus frequency. This occurs when

$$\left(1-\frac{k_t^2\tan\left(\frac{\omega t}{2}\sqrt{\frac{\rho}{c_{33}^D}}\right)}{\frac{\omega t}{2}\sqrt{\frac{\rho}{c_{33}^D}}}\right)=\left(1-\frac{k_t^2\tan\left(\frac{\omega}{4f_p}\right)}{\frac{\omega}{4f_p}}\right)=0 \qquad (0.53)$$

**Table 2.** The equations of the common commercial transducers and the associated parameters of resonance.

| Resonance Mode | Equation | Coefficients |
|---|---|---|
| **Sphere Extensional** | $Y(\omega)=\frac{i\omega 4\pi a^2}{t}\left(\left(1-\mathrm{k}_p^2\right)\boldsymbol{\varepsilon}_{33}^T+\frac{\mathrm{k}_p^2\boldsymbol{\varepsilon}_{33}^T\boldsymbol{\omega}_s^2}{\boldsymbol{\omega}_s^2-\omega^2}\right)$ (Berlincourt et al. 1964) (Tasker et al. 1999) | $\boldsymbol{k}_p^2=\frac{\boldsymbol{\omega}_p^2-\boldsymbol{\omega}_s^2}{\boldsymbol{\omega}_p^2}=\frac{\boldsymbol{d}_{13}^2}{\boldsymbol{\varepsilon}_{33}^T\boldsymbol{s}_C^E}$ $\boldsymbol{f}_s=\frac{1}{2\pi a\sqrt{\rho\mathbf{s}_C^E}}$ with $\mathbf{s}_C^E=\frac{\mathbf{s}_{11}^E+\mathbf{s}_{12}^E}{2}$ |
| **Radial Poled Cylinder** | $\mathbf{Y}(\omega)=\left(\frac{i\omega\boldsymbol{\varepsilon}_{33}^T A}{t}\right)\left[\frac{\alpha_4}{\alpha_1}+\mathbf{k}_{31}^2\left[\frac{\alpha_3^2\tan\left(\frac{KL}{2}\right)}{\alpha_1\alpha_2\left(\frac{KL}{2}\right)}\right]\right]$ (Haskins and Walsh 1957), ( Ebenezer and Sujatha 1997), ( Ebenezer 1996) ( Ebenezer and Abraham 2002) | $\boldsymbol{K}=\omega/\boldsymbol{c}_R,\ \boldsymbol{\Omega}=\omega a/\boldsymbol{c},\ \boldsymbol{c}_R=\boldsymbol{c}\sqrt{\frac{\boldsymbol{\alpha}_2}{\boldsymbol{\alpha}_1}}$ $\boldsymbol{\alpha}_1=1-\left(1-\boldsymbol{\sigma}^2\right)\boldsymbol{\Omega}^2,\boldsymbol{\alpha}_2=1-\boldsymbol{\Omega}^2,\boldsymbol{\alpha}_3=1-\left(1+\boldsymbol{\sigma}\right)\boldsymbol{\Omega}^2$ $\boldsymbol{\alpha}_4=1-\boldsymbol{k}_{31}^2-\left(1+\boldsymbol{\sigma}\right)\left(1-\boldsymbol{\sigma}-2\boldsymbol{k}_{31}^2\right)\boldsymbol{\Omega}^2,$ $\boldsymbol{c}=\sqrt{\frac{1}{\rho\boldsymbol{s}_{11}^E}},,\ \boldsymbol{\sigma}=-\frac{\boldsymbol{s}_{12}^E}{\boldsymbol{s}_{11}^E},\ \boldsymbol{k}_{31}^2=\frac{\boldsymbol{d}_{31}^2}{\boldsymbol{\varepsilon}_{33}^T\boldsymbol{s}_{11}^E}$ |
| **Stack** | $\mathbf{Y}(\omega)=i\omega n\mathbf{C}_0+\frac{2\mathbf{N}^2}{\mathbf{Z}_{ST}}\tanh\left(\frac{n\boldsymbol{\gamma}}{2}\right)$ (Martin 1963, 1964a, 1964b) (Sherrit et al. 2000) | $\boldsymbol{C}_0=\frac{\boldsymbol{\varepsilon}_{33}^T A}{L}(1-\boldsymbol{k}_{33}^2),\ \boldsymbol{N}=\frac{A}{L}\frac{\boldsymbol{d}_{33}}{\boldsymbol{s}_{33}^E},\ \boldsymbol{Z}_{ST}=\left(\boldsymbol{Z}_1\boldsymbol{Z}_2\left(2+\frac{\boldsymbol{Z}_1}{\boldsymbol{Z}_2}\right)\right)^{1/2}$ $\boldsymbol{\gamma}=2\arcsin h\left(\left(\frac{\boldsymbol{Z}_1}{2\boldsymbol{Z}_2}\right)^{1/2}\right),\ \boldsymbol{Z}_1=i\rho\boldsymbol{v}^D A\tan\left(\frac{\omega L}{2\boldsymbol{v}^D}\right)$ $\boldsymbol{Z}_2=\frac{\rho\boldsymbol{v}^D A}{i\sin\left(\frac{\omega L}{\boldsymbol{v}^D}\right)}+\frac{i\boldsymbol{N}^2}{\omega\boldsymbol{C}_0},\ \boldsymbol{v}_D=\frac{1}{\sqrt{\rho\boldsymbol{s}_{33}^D}}$ |
| **Bimorph** | $\boldsymbol{Y}=\frac{i\omega Lw\boldsymbol{\varepsilon}_{33}^T}{2h}\left[1+\boldsymbol{k}_{31}^2\left(\frac{3}{4}\frac{F(\boldsymbol{\Omega}L)}{\boldsymbol{\Omega}L}-1\right)\right]$ $F(\Omega L)=\frac{\cosh(\Omega L)\sin(\Omega L)+\cos(\Omega L)\sinh(\Omega L)}{1+\cos(\Omega L)\cosh(\Omega L)}$ (Smits and Choi 1994) (Smits et al. 1997), (Sherrit et al. 1999) | $\boldsymbol{\Omega}=\left(\frac{\omega^2\rho A_{cs}}{\boldsymbol{EI}}\right)^{1/4}=\sqrt{\frac{\omega}{a}}$ $a^2=\left(\frac{EI}{\rho A_{cs}}\right)=\left(\frac{L^2\omega_s}{R_1^2}\right)^2$ $R_1=1.8751$ $k_{13}=\left(1-\frac{3}{4}\frac{F(x)}{x}\right)^{-1/2}$ |

At the series resonance $\omega=\omega_s$ and equation (0.53) can be written as

$$\frac{\omega_s}{4f_p}=k_t^2\tan\left(\frac{\omega_s}{4f_p}\right), \qquad (0.54)$$

which can be arranged to determine an equation for the electromechanical coupling constant in terms of the series and parallel resonance frequencies of the sample. $k_t$ is then given by

$$k_t^2 = \frac{\pi}{2}\frac{f_s}{f_p}\tan\left[\frac{\pi}{2}\left(1-\frac{f_s}{f_p}\right)\right] \tag{0.55}$$

The value of the clamped permittivity is then found by noting that it is equal to the high frequency permittivity above resonance.

$$\varepsilon_{HF} = \varepsilon_{33}^S \tag{0.56}$$

or by noting that the low frequency permittivity measured below the thickness resonance is of the form

$$\varepsilon_{LF} = \frac{\varepsilon_{33}^S}{1-k_t^2} \tag{0.57}$$

The complex part of the dielectric permittivity is determined by the loss tangent D = tanδ. It is apparent from equations (0.56) and (0.57) that the dielectric loss can only be determined from the high frequency permittivity shown in equation (0.56) since the low frequency loss also has a complex electromechanical coupling term. Only the real part of the electromechanical coupling constant is determined from equation (0.55) and the loss component is assumed to be small. Ignoring the loss component of $k_t$ introduces a significant error in the imaginary component of the clamped permittivity $\varepsilon_{33}^S$ and a smaller error in the real part of the clamped permittivity. The piezoelectric constant governing the thickness mode may be found using the definition of the thickness electromechanical coupling constant

$$h_{33} = k_t\sqrt{\frac{c_{33}^D}{\varepsilon_{33}^S}}\,. \tag{0.58}$$

The other material constants determined from the thickness resonator are

$$e_{33} = h_{33}\varepsilon_{33}^S\,, \tag{0.59}$$

and

$$c_{33}^E = c_{33}^D(1-k_t^2) \tag{0.60}$$

where $e_{33}$ is the piezoelectric charge coefficient, and $c_{33}^E$ is the elastic stiffness at constant electric field. The IEEE Standard equations for analyzing the other modes are summarized in Table 3.

When the spectra displays sideband resonances that obscure the resonance frequencies, another method developed by (Onoe et al. 1963) may be more accurate if the dispersion in the material properties is not significant. This method is usually referred to as the frequency ratio method and relationships between the ratio of the 2nd and fundamental series resonance frequencies and the coupling $k_t$,$k_{33}$, $k_{15}$ have been found and tabularized (Onoe et al. 1963). Polynomials relating the coupling k to polynomials in $r_s = f_{s2}/f_{s1}$ have also been published (Sherrit et al. 1992). In addition to the extensional modes mentioned above, a similar technique for the transverse length thickness mode has been published (Sherrit et al. 1997a).

The radial mode is a special case in that it requires 3 critical frequencies to analyze. The frequency ratio technique (IEEE Standards ANSI/IEEE 176 -1987) for analyzing the radial mode is based on the work of (Meitzler et al. 1973). They defined the series frequency ratio $r_s = f_{s2}/f_{s1}$ in terms of the ratio of the second series resonance frequency to the fundamental series resonance frequency. A relationship was then noted between the series frequency ratio $r_s$ and Poisson's ratio $\sigma^p = -\frac{s_{12}^E}{s_{11}^E}$ and a dimensionless constant η. The relationship was reported in tabular form and used to find the elastic stiffness governing the radial mode $c_{11}^p$ from the equation

**Table 3.** The IEEE Standard on Piezoelectricity (IEEE Standards ANSI/IEEE 176 -1987) equations for analyzing impedance resonances of the thickness, thickness shear, length, and length thickness modes

| | | | | |
|---|---|---|---|---|
| Thickness Extensional Mode | | | | |
| $c_{33}^D = 4\rho t^2 f_p^2$ | $k_t^2 = \frac{\pi}{2}\frac{f_s}{f_p}\tan\left[\frac{\pi}{2}\left(1-\frac{f_s}{f_p}\right)\right]$ | $\varepsilon_{HF} = \varepsilon_{33}^S$ | $\varepsilon_{LF} = \frac{\varepsilon_{33}^S}{1-k_t^2}$ | $h_{33} = k_t\sqrt{\frac{c_{33}^D}{\varepsilon_{33}^S}}$ |
| Thickness Shear Extensional Mode | | | | |
| $c_{55}^D = 4\rho t^2 f_p^2$ | $k_{15}^2 = \frac{\pi}{2}\frac{f_s}{f_p}\tan\left[\frac{\pi}{2}\left(1-\frac{f_s}{f_p}\right)\right]$ | $\varepsilon_{HF} = \varepsilon_{11}^S$ | $\varepsilon_{LF} = \frac{\varepsilon_{11}^S}{1-k_{15}^2}$ | $h_{15} = k_{15}\sqrt{\frac{c_{55}^D}{\varepsilon_{11}^S}}$ |
| Length Extensional Mode | | | | |
| $s_{33}^D = \frac{1}{4\rho l^2 f_p^2}$ | $k_{33}^2 = \frac{\pi}{2}\frac{f_s}{f_p}\tan\left[\frac{\pi}{2}\left(1-\frac{f_s}{f_p}\right)\right]$ | $\varepsilon_{HF} = \varepsilon_{33}^S$ | $\varepsilon_{LF} = \frac{\varepsilon_{33}^S}{1-k_{33}^2} = \varepsilon_{33}^T$ | $d_{33} = k_{33}\sqrt{s_{33}^E\varepsilon_{33}^T}$ |
| Length Thickness Extensional Mode | | | | |
| $s_{11}^E = \frac{1}{4\rho l^2 f_s^2}$ | $\frac{k_{13}^2}{1-k_{13}^2} = \frac{\pi}{2}\frac{f_p}{f_s}\tan\left[\frac{\pi}{2}\left(\frac{f_p}{f_s}-1\right)\right]$ | $\varepsilon_{HF} = \varepsilon_{33}^T(1-k_{13}^2)$ | $\varepsilon_{LF} = \varepsilon_{33}^T$ | $d_{13} = k_{13}\sqrt{s_{11}^E\varepsilon_{33}^T}$ |

$$c_{11}^p = \frac{4\pi^2 f_{S1}^2 a^2 \rho}{\eta^2} \tag{0.61}$$

where $a$ is the sample radius and ρ is the sample density. The standard material constants $s_{11}^E$ and $s_{12}^E$ are then calculated from

$$s_{11}^E = \frac{1}{c_{11}^p(1-(\sigma^p)^2)} \tag{0.62}$$

$$s_{12}^E = \frac{-\sigma^p}{c_{11}^p(1-(\sigma^p)^2)} \tag{0.63}$$

The radial electromechanical coupling constant $k^p$ is determined from an equation of the form.

$$(k^p)^2 = \frac{1-\sigma^p - \mathbf{J}\left(\omega_p a\sqrt{\frac{\rho}{c_{11}^p}}\right)}{2} \tag{0.64}$$

where

$$\mathbf{J}(x) = x\frac{J_0(x)}{J_1(x)} \tag{0.65}$$

In their paper (Meitzler et al. 1973) presented curves at various values of $\sigma^P$ relating the planar coupling factor $k_p$ to $(f_{P1} - f_{S1})/f_{S1}$ where $f_{P1}$ and $f_{S1}$ are the first parallel and series resonance frequencies and the planar coupling factor is defined by

$$(k_p)^2 = \frac{1}{1+\frac{2(k^p)^2}{1+\sigma^p}} \tag{0.66}$$

As the frequency is decreased, the permittivity of a disk resonator, described by radial mode resonance equation in Table 1 can be shown to approach the free permittivity $\varepsilon_{33}^{T}$. The value of the radial permittivity $\varepsilon_{33}^{p}$ is then found using

$$\varepsilon_{33}^{p} = \varepsilon_{33}^{T}(1-k_p^2) \tag{0.67}$$

This technique can be automated (Sherrit et al. 1991) without referring to tables or graphs by using a polynomial to represent the data described by (Meitzler et al. 1973). It was found that the data could be represented accurately by polynomials of the form.

$$\eta = a_0 + a_1 r_s + a_2 r_s^2 + a_3 r_s^3 \tag{0.68}$$

$$\sigma^{p} = b_0 + b_1 r_s + b_2 r_s^2 + b_3 r_s^3 + b_4 r_s^4 \tag{0.69}$$

where the $r_s = f_{s2}/f_{s1}$ is the series resonance ratio. The coefficients of the polynomials are shown in Table 4. The radial electromechanical coupling constant was calculated exactly from equation (0.64) using the series expansion for the zero and first order Bessel function of the first kind using

$$J_0(x) = \sum_{0}^{n=\infty} \frac{(-1)^n}{n!n!}\left[\frac{x}{2}\right]^{2n} \tag{0.70}$$

$$J_1(x) = \sum_{0}^{n=\infty} \frac{(-1)^n}{n!(n+1)!}\left[\frac{x}{2}\right]^{2n+1} \tag{0.71}$$

At resonance the value of the argument of x is x ≈ 2 and x/2 ≈1 which forces the sums in equation (0.70) and (0.71) to converge rapidly. n = 10 is sufficient to calculate $J_0$ and $J_1$ to nine significant figures.

**Table 4.** Coefficients for the polynomials shown in equation (0.68) and (0.69) .

| n | $a_n$ | $b_n$ |
|---|---|---|
| 0 | 11.2924 | 97.527023 |
| 1 | -7.63859 | -126.91730 |
| 2 | 2.13559 | 63.400384 |
| 3 | -.215782 | -14.340444 |
| 4 | | 1.2312109 |

Until now we have discussed the generally accepted methods for determining the real parts of the coupling, elastic and dielectric coefficients from the resonance spectra. A variety of methods have been developed to determine the loss components. These include specific iterative methods (Smits 1976, 1985), (Alemany et al. 1994, 1995), and non-iterative (Sherrit et al. 1991,1992), (Holland and Eernisse 1969), (Du et al. 2003) and general non-linear regression techniques (Kwok et al. 1997), (Tasker et al. 1999) and (Lukacs et al. 1999). When determining the utility of any curve fitting method it is useful to differentiate between accuracy and sensitivity of each of the methods and in this regard it is important to separate the error of the method and the error associated with limited significant figures of the data. The accuracy of the curve fitting method can be deduced by generating spectra of a particular mode from reasonable material coefficients (Holland 1967) and using the method to determine the input material coefficients. The sensitivity of the method can be determined by reducing the number of significant figures and/or adding a random error and then applying the method to determine the error from the input material coefficients. It should also be noted that a model error may be present as is the case when significant dispersion in the material coefficients is present. In this case the impedance equation to which the data is fitted assumes that the coefficients are constant and an error results due to the change of the coefficient over the frequency range of the measurement. Smits' method (Smits 1976, 1985) is a very powerful fitting technique in that it will fit three points of the data exactly in the convergence limit. This means that one can exclude points visually or do multiple analyses using groups of points to average the results. The length thickness LT resonator was covered in depth by Smits in previous work (Smits 1976) so let us look here at the most widely used resonator, the thickness extensional mode.

This method uses three points $Z_0(R_0, X_0, \omega_0)$, $Z_1(R_1, X_1, \omega_1)$, $Z_2(R_2, X_2, \omega_2)$ chosen around resonance. Typically $\omega_0$ corresponds to the frequency of maximum resistance of the spectra. $\omega_1$ and $\omega_2$ are chosen so that they typically correspond to the frequencies of maximum piezoelectric energy in the spectra. Two of the points plus an initial guess for the elastic constant using the method of (Land et al. 1964) or (Sherrit et al. 1992b) are used to calculate the electromechanical coupling constant and the permittivity. Using the coupling constant, permittivity and a third point a revised elastic constant is calculated and the process is repeated until all constants converge. The method evaluates the material constants around a limited region about the resonance and the values are only valid in this region.

The equation governing the resonance of the extensional mode resonators is

$$\mathbf{Z} = (t/i\omega\boldsymbol{\varepsilon}A)\left[1 - \frac{\mathbf{k}^2 \tan(\omega/4\mathbf{f}_p)}{(\omega/4\mathbf{f}_p)}\right] \tag{0.72}$$

where t is the sample thickness and A is the electrode area. All constants shown in bold are complex. For the thickness extensional mode the complex permittivity

$$\boldsymbol{\varepsilon} = \boldsymbol{\varepsilon}_{33}^S \tag{0.73}$$

is the clamped permittivity, The complex electromechanical coupling constant

$$\mathbf{k} = \mathbf{k}_t \tag{0.74}$$

is the complex thickness extensional coupling constant and the complex parallel frequency constant is defined by

$$\left[\mathbf{f}_p\right] = \left[\frac{\mathbf{c}_{33}^D}{4\rho t^2}\right]^{1/2} = \left[\frac{\mathbf{c}_{33}^E}{4\rho\left(1-\mathbf{k}_t^2\right)t^2}\right]^{1/2} \tag{0.75}$$

and is a function of the thickness t, the density ρ and the elastic stiffness

$$\mathbf{c}_{33}^D = \mathbf{c}_{33}^E / (1-\mathbf{k}_t^2) \tag{0.76}$$

The equation for the thickness resonator shown above can be rewritten as

$$\mathbf{Z}(\omega) = \mathbf{a}(\omega)\mathbf{A} - \mathbf{b}(\omega)\mathbf{B} \tag{0.77}$$

where

$$\mathbf{a}(\omega) = \frac{t}{i\omega A} \qquad \mathbf{b}(\omega, \mathbf{f}_p) = \frac{4t\mathbf{f}_p \tan\left(\omega/4\mathbf{f}_p\right)}{i\omega^2 A} \tag{0.78}$$

and

$$\mathbf{A} = \frac{1}{\boldsymbol{\varepsilon}_{33}^S} \qquad \mathbf{B} = \frac{\mathbf{k}_t^2}{\boldsymbol{\varepsilon}_{33}^S} \tag{0.79}$$

Now using the first two points the impedance may be written as two linear equations

$$\begin{bmatrix} \mathbf{Z}_0(R_0, X_0, \omega_0) \\ \mathbf{Z}_1(R_1, X_1, \omega_1) \end{bmatrix} = \begin{bmatrix} \mathbf{a}(\omega_0) & -\mathbf{b}(\omega_0, \mathbf{f}_p) \\ \mathbf{a}(\omega_1) & -\mathbf{b}(\omega_1, \mathbf{f}_p) \end{bmatrix} \begin{bmatrix} \mathbf{A} \\ \mathbf{B} \end{bmatrix} \tag{0.80}$$

The initial guess for $\mathbf{f}_\mathrm{p}$ and the geometry along with the appropriate frequency is used to calculate the values of $\mathbf{a}(\omega)$ and $\mathbf{b}(\omega, \mathbf{f}_p)$ using the identity for the tangent with a complex argument

$$\tan(x+iy) = \frac{\sin 2x + i \sinh 2y}{\cos 2x + \cosh 2y} \tag{0.81}$$

Now, the first set of values of **A** and **B** may be determined by inverting the matrix with complex constants and multiplying by the impedance vector as shown.

$$\begin{bmatrix} \mathbf{A} \\ \mathbf{B} \end{bmatrix} = \begin{bmatrix} \mathbf{a}(\omega_0) & -\mathbf{b}(\omega_0, \mathbf{f}_p) \\ \mathbf{a}(\omega_1) & -\mathbf{b}(\omega_1, \mathbf{f}_p) \end{bmatrix}^{-1} \begin{bmatrix} \mathbf{Z}_0(R_0, X_0, \omega_0) \\ \mathbf{Z}_1(R_1, X_1, \omega_1) \end{bmatrix} \tag{0.82}$$

Using **A** and **B** and the third data point $\mathbf{Z}_2(R_2, X_2, w_2)$ and the redefined equation for the thickness resonance to get

$$\mathbf{b}\left(\omega_2,\mathbf{f}_p\right)=\left(\frac{\mathbf{Aa}\left(\omega_2\right)-\mathbf{Z}_2\left(R_2,X_2,\omega_2\right)}{\mathbf{B}}\right) \tag{0.83}$$

The equation for $\mathbf{b}\left(\omega_2,\mathbf{f}_p\right)$ is a very complicated transcendental function that is difficult to solve: however, one can simplify the analysis, as noted by (Smits 1976), by using the second equation in equation (0.78)

$$\mathbf{b}\left(\omega_2,\mathbf{f}_p\right)=\frac{4t\mathbf{f}_p\tan\left(\omega_2/4\mathbf{f}_p\right)}{i\omega_2^2 A} \tag{0.84}$$

and noting that around resonance the function is dominated by the tangent function. Therefore the tangent function may be isolated as.

$$\tan\left(\frac{\omega_2}{4\mathbf{f}_p^{n+1}}\right)=\left(\frac{\mathbf{Aa}\left(\omega_2\right)-\mathbf{Z}_2\left(R_2,X_2,\omega_2\right)}{\mathbf{B}}\right)\left(\frac{i\omega_2^2 A}{4t\mathbf{f}_p^n}\right) \tag{0.85}$$

or

$$\mathbf{f}_p^{n+1}=\left[\frac{4}{\omega_2}\arctan\left[\left(\frac{\mathbf{Aa}\left(\omega_2\right)-\mathbf{Z}_2\left(R_2,X_2,\omega_2\right)}{\mathbf{B}}\right)\left(\frac{i\omega_2^2 A}{4t\mathbf{f}_p^n}\right)\right]\right]^{-1} \tag{0.86}$$

where $\mathbf{f}_\mathrm{p}^\mathrm{n}$ is the value of $\mathbf{f}_\mathrm{p}$ used to evaluate **A** and **B** initially. The arctan with a complex argument is evaluated using

$$\arctan(x+iy)=\frac{1}{2i}\ln\left[\frac{1-y+ix}{1+y-ix}\right] \tag{0.87}$$

$$\ln(x+iy)=i\arctan\left(\frac{y}{x}\right)+\ln(x^2+y^2)^{1/2} \tag{0.88}$$

by taking special care to evaluate the arctan in the correct quadrant. The process is repeated until the values of **A** , **B** and $\mathbf{f}_\mathrm{p}$ are found to converge. Once **A** , **B** and $\mathbf{f}_\mathrm{p}$ have converged the values of $\boldsymbol{\varepsilon}_{33}^S, \mathbf{k}_t$ *and* $\mathbf{c}_{33}^D$ are found using

$$\boldsymbol{\varepsilon}_{33}^S=\frac{1}{\mathbf{A}} \qquad \mathbf{k}_t=\sqrt{\frac{\mathbf{B}}{\mathbf{A}}} \qquad \mathbf{c}_{33}^D=4\rho t^2\mathbf{f}_p^2 \tag{0.89}$$

and the piezoelectric coefficients $\mathbf{e}_{33}$ and $\mathbf{h}_{33}$ are determined from

$$\mathbf{e}_{33}=\mathbf{k}_t\sqrt{\left(\boldsymbol{\varepsilon}_{33}^S\mathbf{c}_{33}^D\right)} \qquad \mathbf{h}_{33}=\mathbf{k}_t\sqrt{\frac{\mathbf{c}_{33}^D}{\boldsymbol{\varepsilon}_{33}^S}} \tag{0.90}$$

The same approach as shown above can be applied to any of the extensional resonance modes with the substitution of the appropriate material coefficients. Complex methods to evaluate the radial mode have also been published (Sherrit et al. 1991), (Alemany et al. 1995).

### *1.3.3 Determination of the reduced Matrix*

In order to determine the complete matrix for a $C_\infty$ material one requires at least 3 of the resonators shown in Figure 1. For example one could use a thin disk to produce a radial and thickness mode resonance along with a long rod to produce a length extensional resonance and a thin plate poled in the lateral direction to produce a thickness shear resonance to obtain all the dielectric, piezoelectric and elastic coefficients except for $\mathbf{s}_{13}^E$, which has to be calculated using the other known coefficients and the thickness mode coefficients as outlined by Smits (Smits 1976). We first determine $\mathbf{c}_{13}^E$ using

$$\mathbf{c}_{13}^E=\frac{\left(\mathbf{e}_{33}-\mathbf{d}_{33}\mathbf{c}_{33}^E\right)}{2\mathbf{d}_{13}} \quad . \tag{0.91}$$

We then determine $\mathbf{e}_{13}$ using

$$\mathbf{e}_{13} = \frac{\left(\boldsymbol{\varepsilon}_{33}^{T} - \boldsymbol{\varepsilon}_{33}^{S} - \mathbf{d}_{33}\mathbf{e}_{33}\right)}{2\mathbf{d}_{13}}, \tag{0.92}$$

and finally we calculate

$$\mathbf{s}_{13}^{E} = \frac{\left(-\mathbf{c}_{13}^{E}\mathbf{s}_{33}^{E}\mathbf{d}_{13}\right)}{\left(\mathbf{e}_{13} - \mathbf{d}_{33}\mathbf{c}_{13}^{E}\right)} \tag{0.93}$$

It should be remembered that these values are complex and that the sign of the piezoelectric constant is not determined by resonance analysis. The constants must therefore have the proper sign of the piezoelectric coefficients before this calculation is undertaken. Once the calculation is done the full [s,d,ε] matrix can be inverted to get the stiffness constants and other piezoelectric and dielectric constants (ie inverse permittivity and the h or e constant. The value of $\mathbf{s}_{13}^{E}$ can also be determined by inverting the 9x9 [$s^E$,$\varepsilon^T$,d] matrix with an initial guess for the value of $\mathbf{s}_{13}^{E}$ to get the [$c^D$,$\beta^S$,h] matrix. The value of $\mathbf{s}_{13}^{E}$ may be adjusted until the value of $\mathbf{c}_{33}^{D}$ determined from the inverted matrix equaled the value of $\mathbf{c}_{33}^{D}$ determined from the thickness resonance analysis. It has been our experience that the agreement in the real part of $\mathbf{s}_{13}^{E}$ is remarkable considering the different variables that are used to calculate $\mathbf{s}_{13}^{E}$ and the degree of dispersion in the length extensional mode. The calculation proposed by Smits to calculate $\mathbf{s}_{13}^{E}$ uses all the material constants of the thickness mode while the matrix inversion technique uses only the elastic stiffness $\mathbf{c}_{33}^{D}$. The value of the imaginary part of $\mathbf{s}_{13}^{E}$ determined using the two calculation methods can differ by a substantial amount. This is likely due to the combination of error amplification in the calculations due to matrix inversion (subtraction of two numbers of similar magnitudes) and the increased error in the imaginary part of these coefficients. The complete matrices for a variety of piezoelectric materials hard and soft are shown in Table 5. It has been reported recently (Pardo et al. 2007), (Comyn and Tavernor 2001) and (Cao et al. 1998) that inherent clamping in the thickness shear mode gives material coupling and piezoelectric coefficients that are lower by up to 20% from actual values. It has been shown that for the length shear resonator (Aurelle et al. 1994) an aspect ratio of t/W > 6 is required to calculate accurate results.

### *1.3.4 Representing Resonance Curves with Lumped Circuit Parameters*

In the analysis of piezoelectric materials there is another approach to model the impedance of free piezoelectric vibrators that is geometry independent. For transducer applications, it is sometimes more convenient to fit the impedance plots to lumped circuit models in order to predict the electrical behavior of the resonator, as first suggested by Butterworth (Butterworth 1915) and Cady (Cady 1922). This allows one to investigate the interaction of the resonator with the drive circuitry and in

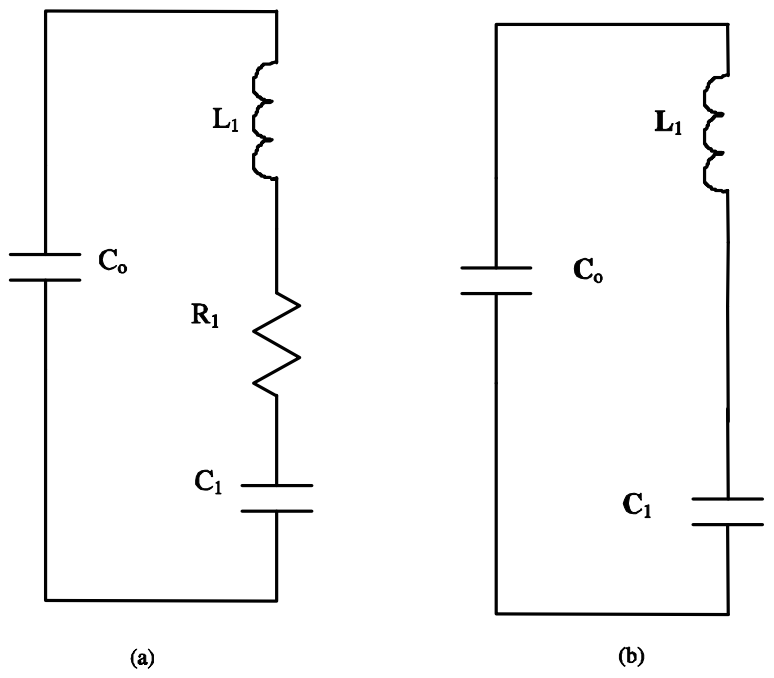

**Fig. 4.** a) The Butterworth Van-Dyke resonator circuit with 4 real parameters and b) the Complex Circuit with 3 complex parameters (6 independent constants).

**Table 5.** The reduced matrix of commercial PZT's and PZT composites including the loss components. The data for $s_{13}^E$ was calculated from the constants of the thickness mode and the other constants shown in the table.

| Material Constant | Motorola 3203HD now CTS (Sherrit et al. 1997d), (Powell et al. 1997) | Ferroperm PZ27** (Pardo et al. 2006) | PZT Channel 5804 (Sherrit et al. 1997b) | EDO EC-65 (Sabat et al. 2007) | EDO EC-69 (Sabat et al. 2007) | Ultrashape Porous PZT UPC 8 – P=21% (Rybianets et al. 2006) | MSI 1-3 PZT-Polymer Composite (30%PZT)*** (Sherrit et al. 1997e) |
|---|---|---|---|---|---|---|---|
| $\mathbf{s}_{11}^E$ (m²/N) x10⁻¹¹ | 1.55 – 0.043i | 1.61 – 0.017i | 1.21 – 0.0045i | 1.71 – 0.02i | 1.12 – 0.01i | 1.96 – 0.0031i | 17.5 – 0.39i |
| $\mathbf{s}_{12}^E$ (m²/N) x10⁻¹¹ | -0.446 +0.018i | -0.59 +0.006i | -0.374 +0.0038i | -0.58 +0.01i | -0.33 +0.01i | -0.637 +0.0023i | -8.10 + 0.39i |
| $\mathbf{s}_{13}^E$ (m²/N) x10⁻¹¹ | -0.819+0.025i | -0.62+0.013i | -0.541 +0.0039i | -0.91 +0.02i | -0.70 +0.02i | -0.542 +0.024i | -2.80 + 0.021i |
| $\mathbf{s}_{33}^E$ (m²/N) x10⁻¹¹ | 1.94 – 0.046i | 1.76 – 0.026i | 1.38 – 0.011i | 1.84 – 0.04i | 1.25 – 0.04i | 3.03 – 0.0013i | 5.19 – 0.13i |
| $\mathbf{s}_{55}^E$ (m²/N) x10⁻¹¹ | 3.92 – 0.13i | 4.08 – 0.18i | 3.70 – 0.010i | 4.6 – 0.56i | 3.25 – 0.16i | 6.25 – 0.23i | 18 – 0.48i |
| $\mathbf{s}_{66}^E$ (m²/N) x10⁻¹¹ | 3.99 – 0.12i | 4.40 – 0.046i | 3.17 – 0.016i | 4.58 – 0.06i | 2.9 – 0.04i | 5.19 – 0.011i | 51.2 – 1.6i |
| $\mathbf{d}_{13}$ (C/N) x10⁻¹² | -294 + 12i | -160 + 3.1i | -108 + 0.37i | -190 – 4.8i | -94.8 - .3i | -59.2 + .052i | -168 + 4.6i |
| $\mathbf{d}_{33}$ (C/N) x10⁻¹² | 584 – 18i | 336 – 7.2i | 219 – 1.3i | 357 – 15i | 201 – 2.2i | 254 – 1.4i | 406 – 15i |
| $\mathbf{d}_{15}$ (C/N) x10⁻¹² | 600 – 30i | 396 – 26i | 415 – 1.1i | 609 – 253i | 422 – 210i | 469 – 26i | 3.5 – 0.15i |
| $\boldsymbol{\varepsilon}_{11}^T$ (F/m) x10⁻⁸ | 2.14 – 0.13i | 1.01 – 0.052i | 1.06 – 0.0091i | 2.05 – 1.1i | 1.43 - .86i | 1.13 - .029i | 0.0095 – 0.0003i |
| $\boldsymbol{\varepsilon}_{33}^T$ (F/m) x10⁻⁸ | 3.01 – 0.081i | 1.37-0.027i | 0.977 – 0.0040i | 1.48 – 0.034i | 0.859 – 0.0024i | 0.644 – 0.00031i | 0.898 – 0.042i |
| ρ (kg/m³) | 7800 | 7700 | 7550 | 7500 | 7500 | 6250 | 3190 |

Full set of constants determined by excluding the free permittivity of the length extensional resonator
** Full set determined from cited paper with reduced $d_{33}$
*** Shear values are less reliable due to the anisotropy of the free permittivity and other shortcomings

some simple resonators the ability to determine velocities of the resonator surfaces. Figure 4a shows the Van Dyke circuit model which is widely used for representing the equivalent circuit of a piezo-electric vibrator (Martin 1954), (Terunuma and Nishigaki 1983) and whose use is recommended by the IEEE standard on piezoelectricity (IEEE Standards ANSI/IEEE 176 -1987).

Figure 4a shows that the Van Dyke model uses four real circuit parameters, $C_0$, $C_1$, $L_1$ and $R_1$, to represent the impedance of a free-standing piezoelectric resonator around resonance. However, equation (0.50) shows that six material constants (3 complex constants) are needed to describe the resonance completely when losses are significant. Thus the Van Dyke model, with only four independent constants, cannot accurately represent the functional relationship shown in equation (0.50), particularly for materials with significant losses. Other authors have tried to generalize the Van Dyke model by including resistive elements in parallel or in series with the other reactive parameters but, as Von Hippel (Von Hippel 1967) has pointed out, representing the losses of a capacitor or inductor by adding a frequency independent resistor in parallel with them is less general than is representing these losses by the use of complex circuit components. Representing the loss as a resistive component leads to the ratio of the loss current to the charging current being a function of the form 1/(ωRC). This form assumes that the conduction term is dominated by the migration of charge carriers. This assumption is not always valid since the electrical loss results from all energy consuming processes which include, amongst others, the phase lag of the dielectric displacement due to electronic, ionic and domain motions. A model that takes this in to account is the complex circuit model shown in Figure 4b derived from the lossless model of Butterworth (Butterworth 1915) and Cady (Cady 1922). In the thickness mode, the parameters $\mathbf{C}_0$, $\mathbf{C}_1$, $\mathbf{L}_1$ are defined as complex and the values can be calculated from the complex frequency data ($\boldsymbol{Y}$=iω$\boldsymbol{C}_{LF}$, $\boldsymbol{f}_s$, $\boldsymbol{f}_p$) using equations (0.94) to (0.96) below or the complex material coefficients ($\boldsymbol{c}_{33}^D, \boldsymbol{\varepsilon}_{33}^S, \boldsymbol{k}_t$) using equations (0.97) - (0.99).

$$\boldsymbol{C}_{LF} = \boldsymbol{C}_0 + \boldsymbol{C}_1 \tag{0.94}$$

$$\boldsymbol{f}_p = \frac{1}{2\pi\sqrt{\mathbf{L}_1 \frac{\mathbf{C}_1\mathbf{C}_0}{\mathbf{C}_1+\mathbf{C}_0}}} \tag{0.95}$$

$$f_s = \frac{1}{2\pi\sqrt{\mathbf{L}_1\mathbf{C}_1}} \tag{0.96}$$

$$C_{LF} = \frac{\varepsilon_{33}^{S} A}{t\left(1-k_t^2\right)} \tag{0.97}$$

$$f_p = \left[\frac{\mathbf{c}_{33}^{E}}{4\left(1-\mathbf{k}_t^2\right)\rho t^2}\right]^{1/2} = \left[\frac{\mathbf{c}_{33}^{D}}{4\rho t^2}\right]^{1/2} \tag{0.98}$$

$$f_s = \frac{2}{\pi}\mathbf{k}_t^2 f_p \cot\left[\frac{\pi}{2}\frac{f_p - f_s}{f_p}\right] \tag{0.99}$$

The determination of the complex circuit parameters from the constants of the data ($Y$=i$\omega C_{LF}$, $f_s$, $f_p$ ) is straight forward but determining the circuit parameters from the complex material constants first requires determining $f_s$ from a transcendental function (Sherrit et al. 1997c). Although we have shown a detailed calculation for the thickness mode resonator, this complex circuit can be extended to all the other standard modes (Sherrit 1997, Sherrit et al. 1997c). In addition the complex circuit model can be collapsed into the Butterworth Van Dyke model by calculating the loss components at the resonance frequency and transforming the losses to the motional branch (Sherrit et al. 1997c).

### *1.3.5 One Dimensional Network Models*

Analytical solutions to the wave equation in piezoelectric materials can be quite cumbersome to derive from first principles in all but a few cases. Mason (Mason 1948, 1958) was able to show that for one-dimensional analysis most of the difficulties in deriving the solutions could be overcome by borrowing from network theory. He presented an exact equivalent circuit that separated the piezoelectric material into an electrical port and two acoustic ports through the use of an ideal electromechanical transformer as shown in Figure 5. The model has been widely used for free and mass loaded resonators (Berlincourt et al. 1964) transient response (Redwood 1961), material constant determination (Saitoh et al. 1985), and a host of other applications (Katz 1959). One of the perceived problems with the model is that it required a negative capacitance at the electrical port although Redwood (Redwood 1961) showed that this capacitance could be transformed to the acoustic side of the transformer and treated as a length of the acoustic line. In an effort to remove circuit elements between the top of the transformer and the node of the acoustic transmission line**,** Krimholtz, Leedom and Matthae (Krimholtz et al. 1970) published an alternative equivalent circuit as shown in Figure 6. The model is commonly referred to as the KLM model and has been used extensively in the medical imaging community in an effort to design high frequency transducers (Zipparo et al. 1997), (Foster et al. 1991) multi-layers (Zhang et al. 1997), and arrays (Goldberg et al. 1997).

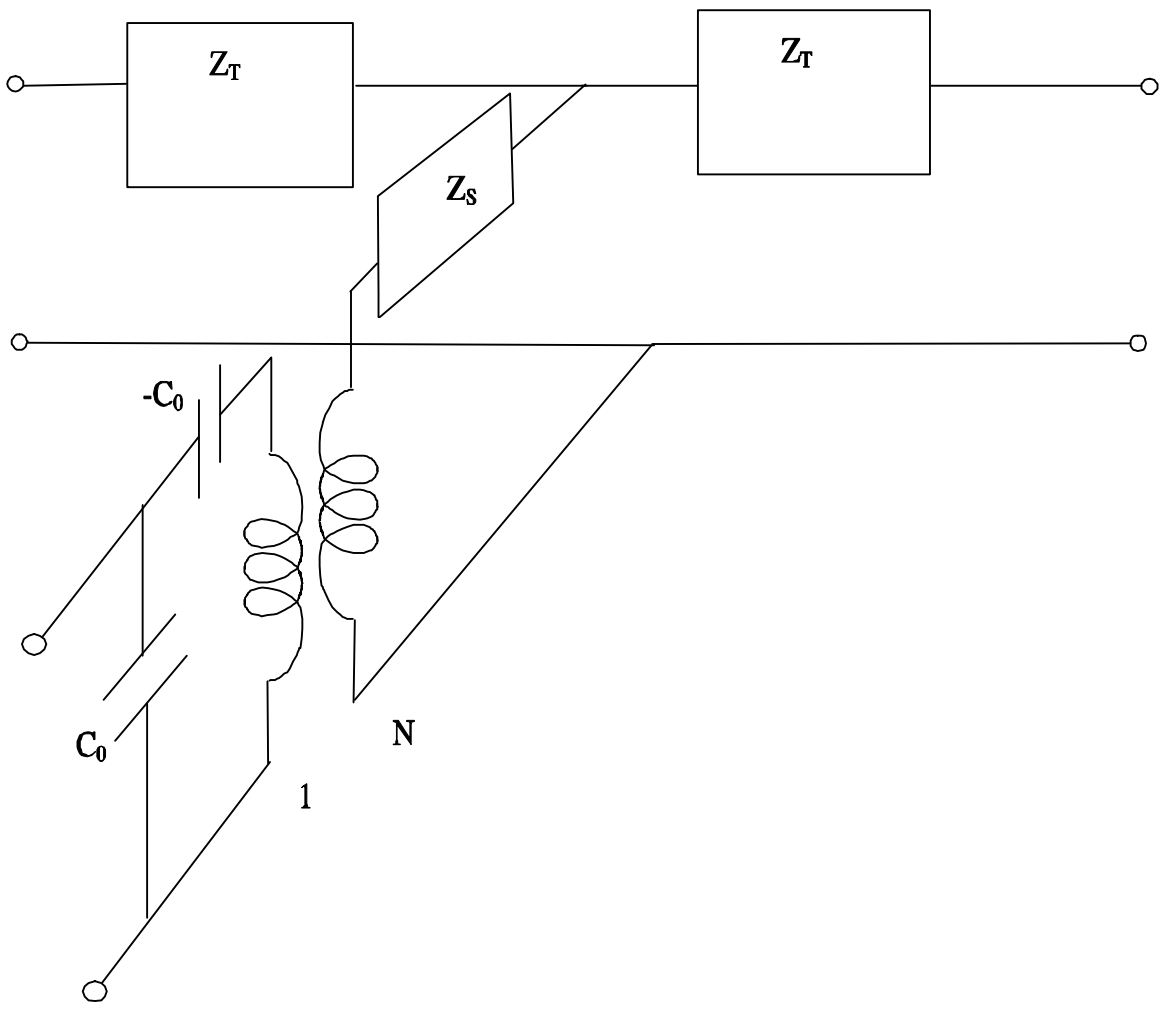

**Fig. 5.** Mason's network model for the piezoelectric resonator in the thickness mode.

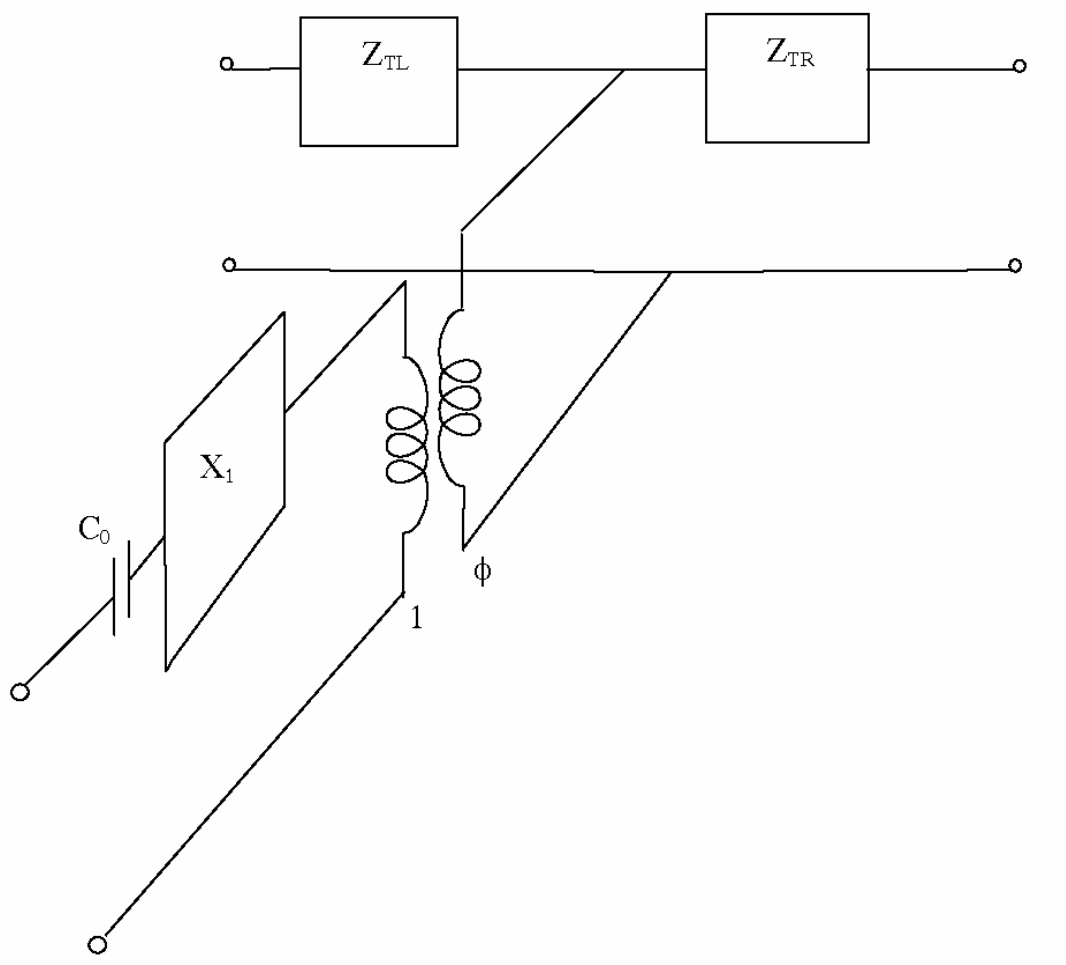

**Fig. 6.** Krimholtz, Leedom and Matthae (Krimholtz et al. 1970) (KLM) equivalent circuit used for transducer modeling.

In the following sections we present the circuit parameter for the KLM and Mason's equivalent circuit for the case where the piezoelectric, dielectric and elastic constants are represented by complex quantities to account for intrinsic loss in the material. The parameters of each model are shown in Table 6. When the acoustic ports are shorted, each of these models has been shown to produce identical impedance spectra to the free resonator equation (0.50) when the loss is applied consistently (Sherrit et al. 1999b). In Mason's and the KLM model the center "tap" is the electrical port where voltage is applied or produced and the left and right "taps" are the acoustic ports and correspond to the faces of the transducer that either transmit or receive a stress. On the electrical side of the transformer the voltages and currents are treated as normal circuit elements ($V = Z_eI$) where $Z_e$ is an electrical impedance. A voltage that is transformed across the transformer to the acoustic side becomes a force. On the acoustic side the force is related to velocity $v$ through $F = Z_a v$ where $Z_a$ is the specific acoustic impedance ($Z_a = \rho v A$). Although the KLM model appears to be easier to deal with in terms of circuit manipulation of series and parallel combinations of piezoelectric elements and acoustic layers, the turns ratio of the transformer is no longer frequency independent. In many cases Mason's equivalent circuit will suffice when the number of piezoelectric elements or acoustic layers are not that large.

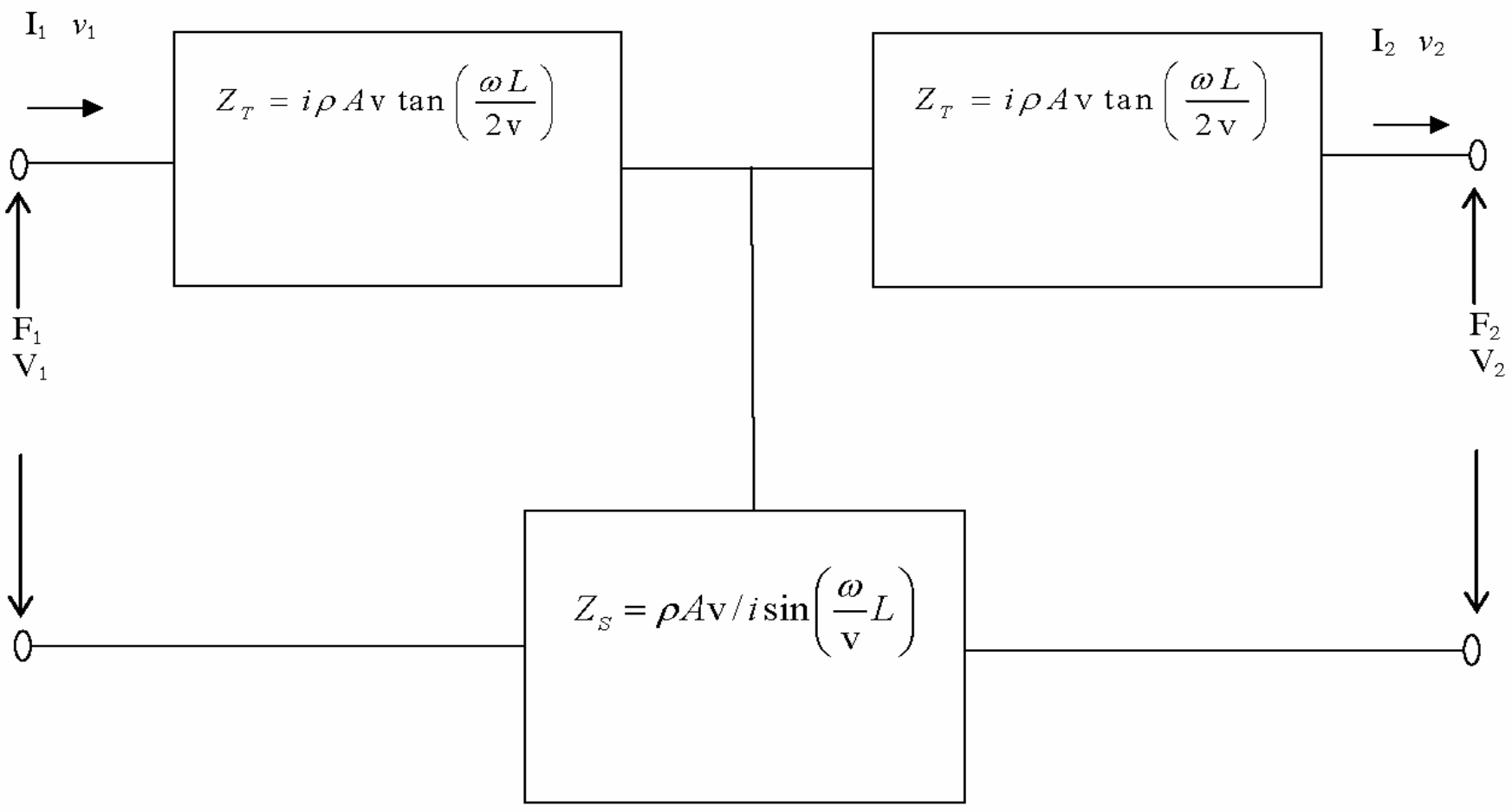

**Fig. 7.** Network representation of the one-dimensional solution to the wave equation for an extensional mode in a plate. The boundary conditions as shown are open. Electrical Analogs are Voltage = Force, Current = Velocity and Specific acoustic impedance is analogous to the electrical impedance. Losses can be accounted for by allowing the velocity to be complex $\text{v}^D=(c^D/\rho)^{1/2}$, $A$= area, $L$ = plate thickness.

As mentioned above, these models can be used to determine the free resonator equation although their primary utility is in the modeling of composite resonators by the addition of acoustic layers (Sherrit et al. 2000). The network equivalent of an acoustic layer in the thickness mode is shown in Figure 7. If this layer network is attached to one of the faces of the piezoelectric as is shown in Figure 8 we have the one dimensional solution of a free standing piezoelectric attached to a substrate. These 1D models were found (Sherrit et al. 1999b) to produce identical results to the analytical solution of the wave equation and the linear equations of a piezoelectric material in a composite piezoelectric/substrate transducer derived by Lukacs et al. (Lukacs et al. 1999). For a simple thickness mode transducer these models can be built up layer by layer to include backing, matching layers and tuning circuit elements and even the medium that the transducer is transmitting or receiving acoustic energy from. These models have also been developed for other standard resonator or transducer geometries (Berlincourt et al. 1964) and the models shown in Figure 5 and Figure 6 can be used to describe the length extensional and thickness shear extensional modes directly with a change in material coefficients. The length thickness mode has also been published along with the ring extensional resonance model (Berlincourt et al. 1964). These models offer a powerful tool in evaluating the frequency response of a transducer and they allow for the calculation of the stresses and velocities of the surfaces directly from the material properties of the piezoelectric, acoustic layers and transmission medium. An example of impedance curves for a piezoelectric with a stainless steel backing and an epoxy backing generated using Mason's and the KLM equivalent circuits are shown in Figure 9. The data used to generate these curves is shown in Table7. Although it is well known that the complex permittivity accounts for the leakage current (Von Hippel 1967) and that the complex stiffness or compliance accounts for the attenuation (McSkimin 1964), a complex turns ratio is a relatively new concept; however, it can be shown to produce a phase shift between the input electrical signal and the generated stress in the piezoelectric.

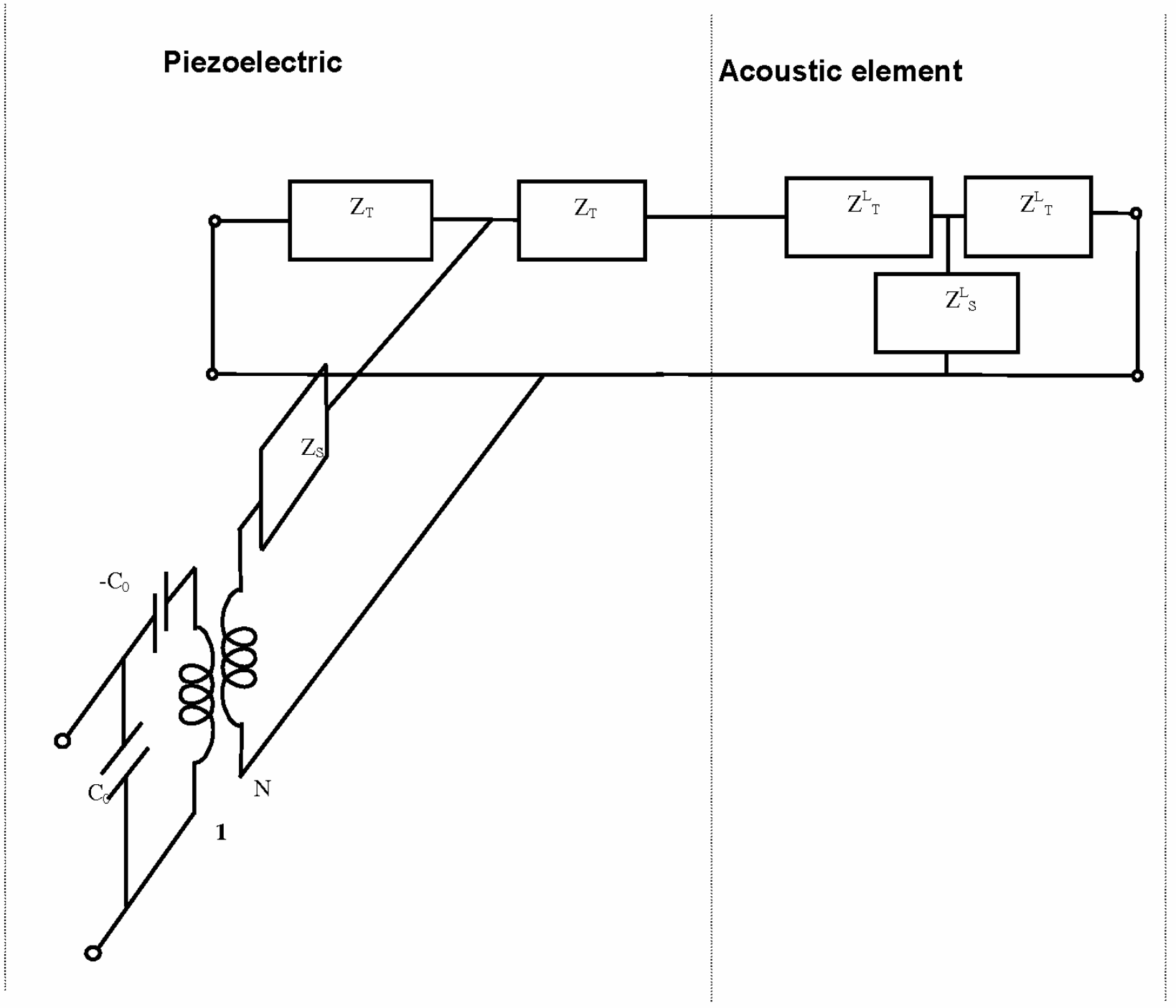

**Fig. 8.** The equivalent circuit representation of the piezoelectric resonator attached to an acoustic element. The mechanical boundary conditions on both exposed surfaces of the element and piezoelectric are unclamped (short circuit on the acoustic port).

To calculate the impedance of the transducer in the case of Mason's equivalent circuit one simply determines the sum of the parallel and series impedances in the network on both the left and right

**Table 6.** The complex material constants and the KLM and Masons parameters

**Mason's Model**

$$\mathbf{C}_0 = \frac{\boldsymbol{\varepsilon}_{33}^S A}{t} \qquad \mathbf{N} = \mathbf{C}_0 \boldsymbol{h}_{33}$$

$$\boldsymbol{Z}_0 = \rho\, A \mathbf{v}^D = A\sqrt{\rho\, \mathbf{c}_{33}^D} \qquad \Gamma = \frac{\omega}{\mathbf{v}^D} = \omega\sqrt{\frac{\rho}{\boldsymbol{c}_{33}^D}}$$

$$Z_T = i\mathbf{Z}_0 \tan\left(\Gamma t/2\right) \qquad Z_S = -i\mathbf{Z}_0 \csc\left(\Gamma t\right)$$

**KLM Model**

$$\mathbf{C}_0 = \frac{\boldsymbol{\varepsilon}_{33}^S A}{t}$$

$$\boldsymbol{Z}_0 = \rho\, A \mathbf{v}^D = A\sqrt{\rho\, \mathbf{c}_{33}^D} \qquad \mathbf{M} = \frac{\boldsymbol{h}_{33}}{\omega \boldsymbol{Z_0}}$$

$$\Gamma = \frac{\omega}{\mathbf{v}^D} = \omega\sqrt{\frac{\rho}{\boldsymbol{c}_{33}^D}} \qquad X_1 = i\mathbf{Z}_0\mathbf{M}^2 \sin(\Gamma t) \qquad Z_{TL} = \mathbf{Z}_0\left[\frac{\mathbf{Z}_L \cos(\Gamma t/2) + i\mathbf{Z}_0 \sin(\Gamma t/2)}{\mathbf{Z}_0 \cos(\Gamma t/2) + i\mathbf{Z}_L \sin(\Gamma t/2)}\right]$$

$$Z_{TR} = \mathbf{Z}_0\left[\frac{\mathbf{Z}_R \cos(\Gamma t/2) + i\mathbf{Z}_0 \sin(\Gamma t/2)}{\mathbf{Z}_0 \cos(\Gamma t/2) + i\mathbf{Z}_R \sin(\Gamma t/2)}\right] \qquad \phi = \frac{1}{2\mathbf{M}}\csc(\Gamma t/2)$$

$Z_L$, $Z_R$ =load impedance on left and right acoustic ports

**Table 7.** The material parameters of the piezoelectric, stainless steel and epoxy layer required to generate the curves using Mason's and the KLM model data shown in Figure 9.

| Material Constants and Geometry of Piezoelectric Material (Motorola 3203HD) |
|---|
| $\rho$ = 7800 kg/m$^3$ t = 0.001 m Diameter = 0.015 m |
| $\mathbf{c}_{33}^D$ (x 10$^{11}$ N/m$^2$) = 1.77 (1 + 0.023i) |
| $\boldsymbol{\varepsilon}_{33}^S$ (x 10$^{-8}$ F/m) = 1.06 (1 - 0.053i) |
| $\boldsymbol{h}_{33}$ (x 10$^9$ V/m) = 2.19 (1 + 0.029i) |
| $k_t$ = 0.536 ( 1 - 0.005i) |
| $C_0$ (nF) = 1.87 (1 – 0.053i) |
| N (C/m) = 4.11 (1 - 0.024i) |
| $\mathbf{v}_D$ (m/s) = 4674 (1 + 0.012i) |
| $\Gamma/\omega$ (x10$^{-4}$ s/m) = 2.10 (1 – 0.012i) |
| M$\omega$ (x10$^5$ Vs/mkg) = 3.33 (1 + 0.017i) |
| Acoustic Layer Properties |
| t = 0.001 m Diameter = 0.015 m |
| Epoxy |
| $\rho$(kg/m$^3$) = 1100 |
| $\mathbf{c}_{33}^D$ (x 10$^9$ N/m$^2$) = 5.3 (1 + 0.1i) |
| $\mathbf{v}_D$ (m/s) = 2200 (1 + 0.05i) |
| $\Gamma/\omega$ (x10$^{-4}$ s/m) = 4.53 (1 – 0.05i) |
| Stainless Steel |
| $\rho$(kg/m$^3$) = 7890 |
| $\mathbf{c}_{33}^D$ (x 10$^{11}$ N/m$^2$) = 2.645 (1 + 0.002i) |
| $\mathbf{v}_D$ (m/s) = 5790 (1 + 0.001i) |
| $\Gamma/\omega$ (x10$^{-4}$ s/m) = 1.727 (1 – 0.001i) |

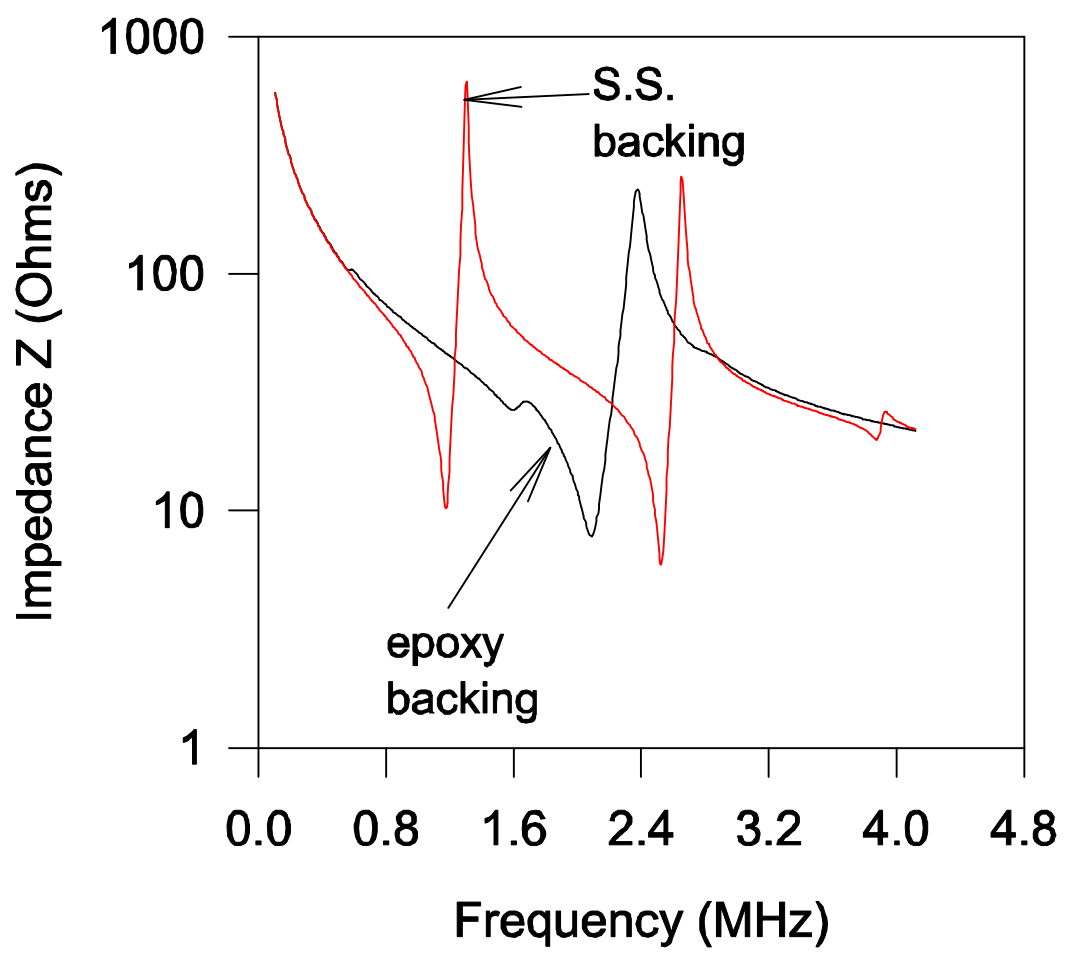

**Fig. 9.** A comparison of the analytical solution, Mason's and the KLM model with stainless steel and epoxy backing layers. The data from each model were found to overlap.

acoustic ports and determines the impedance of the parallel combination. This sum is added to the piezoelectric series impedance $Z_S$ and transformed through the transformer and this result is added to the impedance of the negative complex capacitance. The total impedance is therefore this result added in parallel with the impedance of the positive complex capacitance $C_0$. A similar process is used in the KLM model. The current is used to determine the velocity of a face of the transducer and its displacement under harmonic excitation is simply the velocity divided by iω.

### *1.3.6 Dispersion in Material Coefficients*

All materials have dispersion in the material properties in some frequency region. For example the classic dielectric (Kittel 1976) can have polarization contributions due to the electrons (displacements in the electron shells), ions (core displacements), and molecular dipoles contributions. In piezoelectric ceramics there are also domain contributions. Each of these polarization mechanisms have an associated inertia which causes the net polarization to decrease in plateaus as a function of frequency with the "more inertial" (larger structures) mechanisms clamping out first. Similar reasoning can be applied to the elastic response and domain pinning, grain boundaries, dislocations and impurities can all contribute to the dispersion in the elastic loss (Mason 1950), (Sherrit and Mukherjee 1998a). The piezoelectric coefficient also has a frequency dependence that can be measured at reasonable frequencies (Mukherjee et al. 2001). The technique proposed by Smits (Smits 1976) to analyze the material constants can be extended to fit the impedance equation to higher-order resonances. In order to use Smits' technique to analyze higher-order resonances one has to ensure that the correct quadrant is used to evaluate the function for the correct resonance order. This translates to adding a $\delta$ term to the real component of the arctan function where δ is a function of the resonance order $n$ of the form

$$\delta = \frac{(2n-1)\pi}{2}, \qquad (0.100)$$

where the resonance order is $n$=1 for the fundamental resonance, ($n$=2 for the second resonance, etc.). The material constants can then be analyzed at the fundamental and higher-order resonances and the material constants at each frequency over the frequency range of the spectrum can be determined. The result is a set of three (elastic, dielectric, piezoelectric) complex constants determined at discrete frequencies bands. If the dispersion mechanism is known, the results can then be fit to the dispersion model (e.g., viscoelastic, Debye, etc.). If the specific mechanism is as yet unknown these constants may then be fit to a general polynomial to determine a functional form for the dispersion. The piezoelectric, dielectric and elastic constant then can be written as polynomials of the form

$$Re\, x(f) = a_0 + a_1 f + a_2 f^2 \ldots a_n f^n \qquad (0.101)$$

$$Im\, x(f) = b_0 + b_1 f + b_2 f^2 \ldots b_n f^n, \qquad (0.102)$$

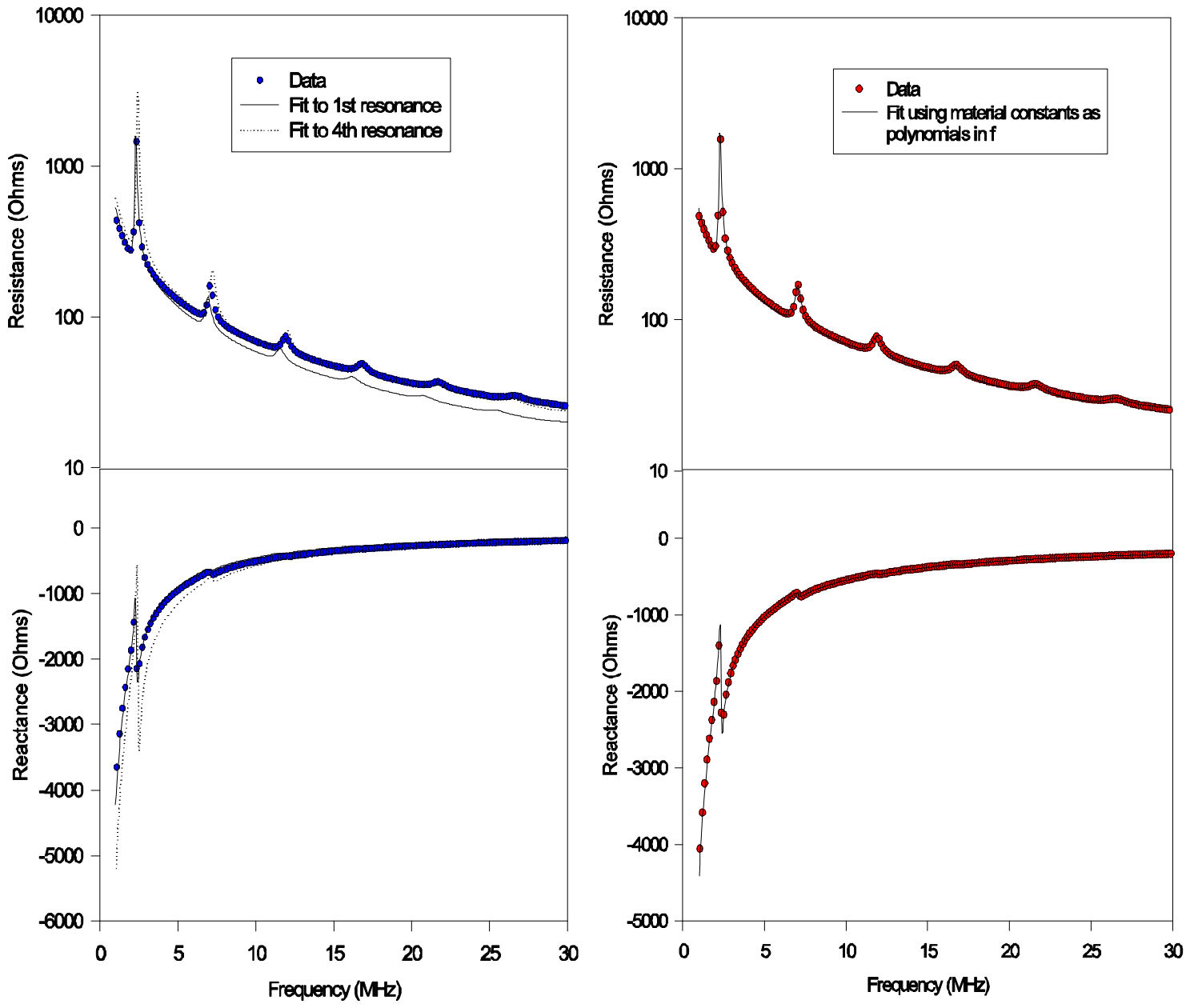

**Fig. 10.** The impedance spectra of a PVDF-TRFE copolymer sample and the fit to the spectra using Smits' technique about the first and 4$^{th}$ resonance. The curves on the right show the fit to the data using the frequency response of the thickness material parameters shown in Table8. The coefficients of the polynomials are shown in Table 8.

where $x$ may be the piezoelectric coefficient ($d$, $g$, $e$, $h$), the elastic constant ($s$ or $c$) or the dielectric constant (ε or β). The maximum allowable value of $n$ in the polynomial equals the number of resonance peaks that have been analyzed so that $n_{\max} = n$ the largest order of the resonance analyzed. The fit of the material constants to the polynomial in $f$ is arbitrary and a polynomial in log $f$ or other functional forms of the dispersion can be used. This approach is valid only for small signal analysis. Driving at large signals can introduce non-linearities and harmonics that should not be included in impedance measurements by definition. An example of this approach is shown in the thickness resonance of a PVDF-TRFE copolymer sample shown in Figure 10. The curves on the left side of the figure show the data and the fit to the 1$^{st}$ and 4$^{th}$ resonance peaks. The dispersion is clearly evident for each curve as you move away from each fit peak. Each of the resonance peaks was analyzed using Smits' method and the results for two separate samples are shown in Figure 11. The coefficients of the polynomials for the 3 complex material constants are listed in Table 8. The frequency dependence of the thickness mode material constants was then applied to the thickness impedance equation and the resultant spectra are shown on the right side of Figure 10. The fit to the impedance data is seen to be excellent over the entire frequency range.

**Table 8.** The coefficients of the polynomials for a PVDF TRFE copolymer sample determined by fitting the data shown in Figure 11 to a generalized polynomial

| Constant | 0th term | 1st term | 2nd term | 3rd term |
|---|---|---|---|---|
| Re $\mathbf{c}_{33}^{D}$ (N/m$^2$) | 8.994x10$^9$ | 116.1 | -4.894x10$^{-6}$ | 7.677x10$^{-14}$ |
| Im $\mathbf{c}_{33}^{D}$ (N/m$^2$) | 5.298x10$^8$ | 1.338 | 1.144x10$^{-7}$ | |
| Re $\boldsymbol{\varepsilon}_{33}^{S}$ (F/m) | 4.793x10$^{-11}$ | -1.283x10$^{-18}$ | 5.686x10$^{-26}$ | -9.253x10$^{-34}$ |
| Im $\boldsymbol{\varepsilon}_{33}^{S}$ (F/m) | -6.106x10$^{-12}$ | 1.382x10$^{-19}$ | -2.753x10$^{-27}$ | |
| Re $k_t$ (#) | 0.2201 | 3.920x10$^{-9}$ | 6.191x10$^{-17}$ | |
| Im $k_t$ (#) | 0.01153 | -1.268x10$^{-9}$ | 4.960x10$^{-17}$ | |

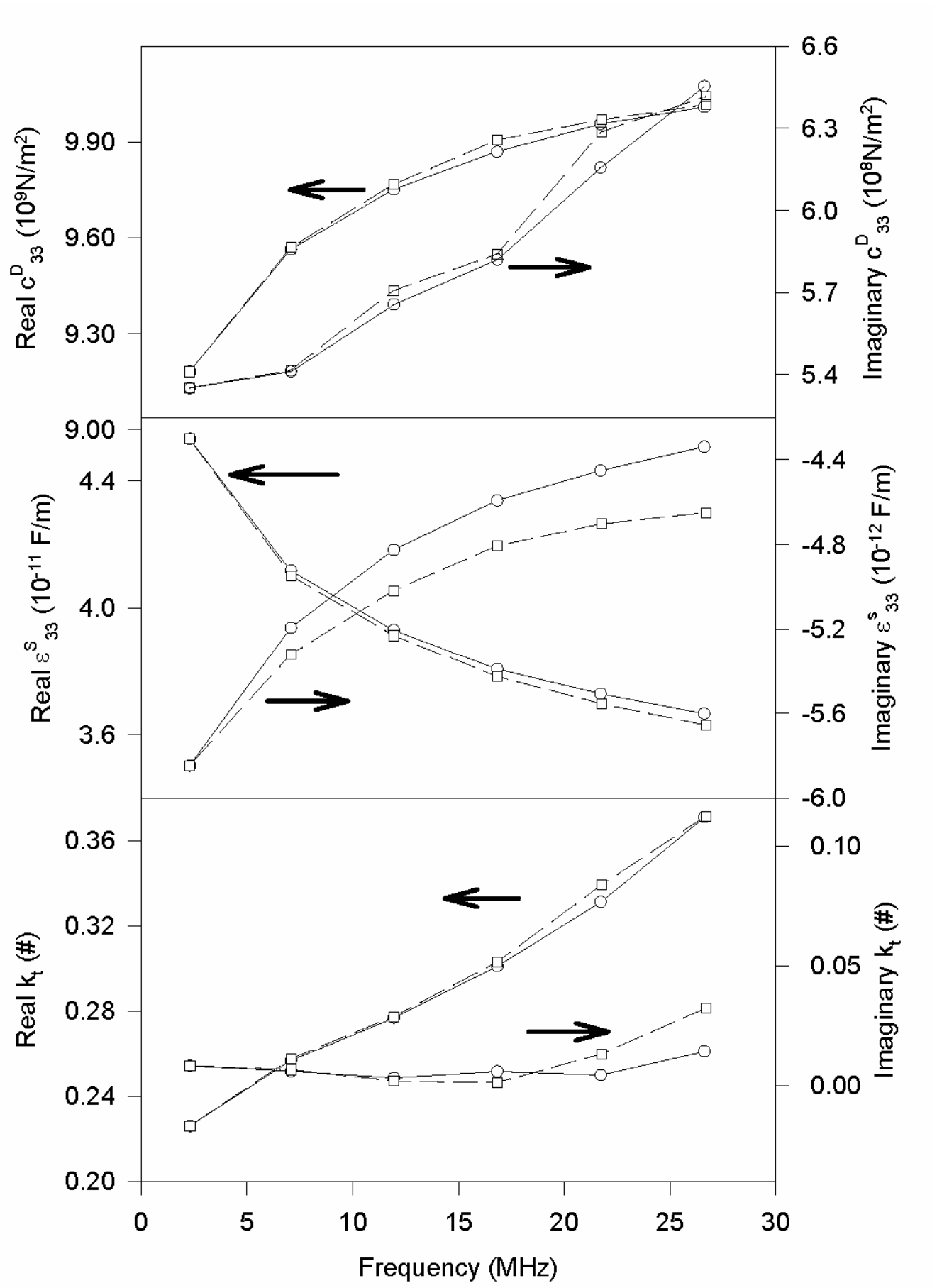

**Fig. 11.** The thickness material constants measured on two different PVDF-TRFE copolymer samples as a function of frequency determined from Smits' method.

## 1.4 Non-linearities

### *1.4.1 Background*

The linear model of piezoelectricity cannot explain all the observed properties of piezoelectric materials and nonlinear effects have been reported by many authors. (Berlincourt and Krueger 1959), (Woolett and Leblanc 1973), (Hall 2001), (Mukherjee et al 2001), (Gonnard 2002) and (Damjanovic and Robert 2002) have looked at various aspects of the nonlinear behavior of piezoelectric materials as a function of applied fields, stress, time, frequency and temperature. Krueger (Krueger 1954, 1967, 1968a, 1968b), (Brown and McMahon 1962, 1964), (Woolett and Leblanc 1973), (Cao and Evans 1993), (Audigier et al. 1994) and (Yang et al. 2000a, 2000b) have studied the stress dependence of the material properties of piezoelectric ceramics. (Land et al, 1964), (Gdula 1968), (Viera 1986), (Hall and Stevenson 1999) and (Masys et al. 2003) have reported on the dependence of various material properties on the applied AC field. (Gonnard and Petit 2002), (Yang et al. 2000b) and (Masys et al. 2003) have investigated the effects of a DC bias field on the piezoelectric properties and (Sabat and Mukherjee 2007) have studied the variation of the material constants of piezoelectric ceramics as a function of temperature. The time dependence of the piezoelectric response after a DC stress step has been studied by (Sherrit et al. 1992c), who used their observations to find average activation energies for non-180° domain wall motions in piezoelectric ceramics. Most if not all the useful piezoelectric ceramics are ferroelectric and their polarization-electric field and strain-electric field relationships become non-linear to reflect the re-orientation of polarization by the electric field and the coupling of the polarization to the strain. Similarly the re-orientation of domains due to stress introduces non-linearities in the stress-strain relationship. If the linear constitutive equations are used to define the material coefficients of the material, then these coefficients themselves are a function of applied fields and stresses as well as of the temperature of the material.

It is useful to look at a generalized non-linear response in the time and frequency domains in order to develop a metric to quantify a non-linear response. An example of a generic non-linear response is shown in Figure 12. In this case it represents a lossy dielectric material driven to the beginning of saturation. The curve was generated using linear and cubic field dependencies for the electric displacement. Assuming an isotropic dielectric response we can express the applied electric field and the resultant electric displacement as

$$\begin{aligned} E(t) &= E_0 \cos(\omega t) \\ D(t) &= \boldsymbol{\varepsilon} E(t) + \boldsymbol{\kappa}\left(E(t)\right)^3 \end{aligned} \tag{0.103}$$

where both $\boldsymbol{\varepsilon}$ and $\boldsymbol{\kappa}$ are complex constants. In order to calculate the time response we need to use the power formulas for the cosine function. These are for order up to n=5;

$$\begin{aligned} \cos(\omega t) &= \cos(\omega t) \\ \cos(\omega t)^2 &= 1/2 + \cos(2\omega t)/2 \\ \cos(\omega t)^3 &= 3/4\cos(\omega t) + \cos(3\omega t)/4 \\ \cos(\omega t)^4 &= 3/8 + \cos(2\omega t)/2 + \cos(4\omega t)/8 \\ \cos(\omega t)^5 &= 5/8\cos(\omega t) + 15\cos(3\omega t)/16 + \cos(5\omega t)/16 \end{aligned} \tag{0.104}$$

Now let us rewrite $\varepsilon$ and $\kappa$ in polar coordinates with a magnitude $|\varepsilon| = 2\text{x}10^{-8}$ F/m and $|\kappa| = -5\text{x}10^{-22}$ Cm/V$^3$ and angle $\phi_\varepsilon = -\pi/30$ $\phi_\kappa = -\pi/18$ and apply these coefficients as shown in the second equation of equations (0.103). The time response of the electric displacement can therefore be written as

$$D(t) = |\boldsymbol{\varepsilon}| E_o \cos(\omega t + \phi_\varepsilon) + |\kappa| E_o^3 \left(3/4\cos(\omega t + \phi_\kappa) + \cos(3\omega t + \phi_\kappa)/4\right) \tag{0.105}$$

and is shown in the top left curve of Figure 12. The frequency component determined from a fast Fourier transform of the time curve shows linear and cubic terms in both the real and imaginary Fourier coefficients. The projection of the D(t) and E(t) curves on the D-E plane is shown in the bottom curve of Figure 12. This particular curve is both lossy and non-linear and the complex shape; saturation, hysteresis and pinching at the turn around points can all be quantified by two complex coefficients. These coefficients can be determined directly from the fast Fourier transform data (Leary and Pilgrim 1998) by adding the linear terms in equation (0.105) and the cubic terms and comparing there amplitudes to the Fourier sum. Although we have used the example of a linear and cubic field dependence the nonlinear response can be generalized to deviations from linearity for the various Y =f(X) curves such as the stress-strain (T-S), electric displacement-stress (D-T) and the strain–field (S-E) curves. Under a sinusoidal excitation each of these responses produces a shifted sinusoid (lossy) or a shifted sinusoidal response with higher order harmonics (lossy and non-linear). The projection of which appears as a non-linear hysteretic response in the X, Y(X) plane.

It is important to understand the different ways a non-linear curve may be quantified when the data is viewed in the X, Y(X) plane and the interesting features of the data that may be present when irreversibility is present. Consider the idealized nonlinear lossy curve representing a ferroelectric/piezoelectric ceramic as is shown in Figure 13. When a very small AC electric field or stress is applied to a piezoelectric ceramic, the measured electric displacement or strain is small, the response is linear, non-linearities are not observed and losses are constant. The slope of the line for this data would be the value of the piezoelectric constant however it is worth noting that we also have the average slope which depends on the maximum field applied, the differential slope which depends on the field level and the maximum field applied and the differential measurement (small signal at some set bias field). (Zhang et al. 1988) have shown that each piezoelectric material has a plateau region where the dielectric constant and the piezoelectric constant are independent of field and they attributed this field independence to reversible domain changes. In the case of the resonance measurements, the applied AC electric fields are small and within the plateau region and the coefficients that are measured correspond to the reversible contribution only. At larger values of the applied electric field or of applied stress, irreversible domain changes occur and make an additional contribution to the piezoelectric and dielectric effects. The average slope of the hysteresis curve is now greater than the slope of the linear response at low fields and this average slope can be measured and as a function of the maximum value of the applied field. Ideally however one would like to separate out the linear and non-linear terms and this can be done by determining the frequency components of time base signals.

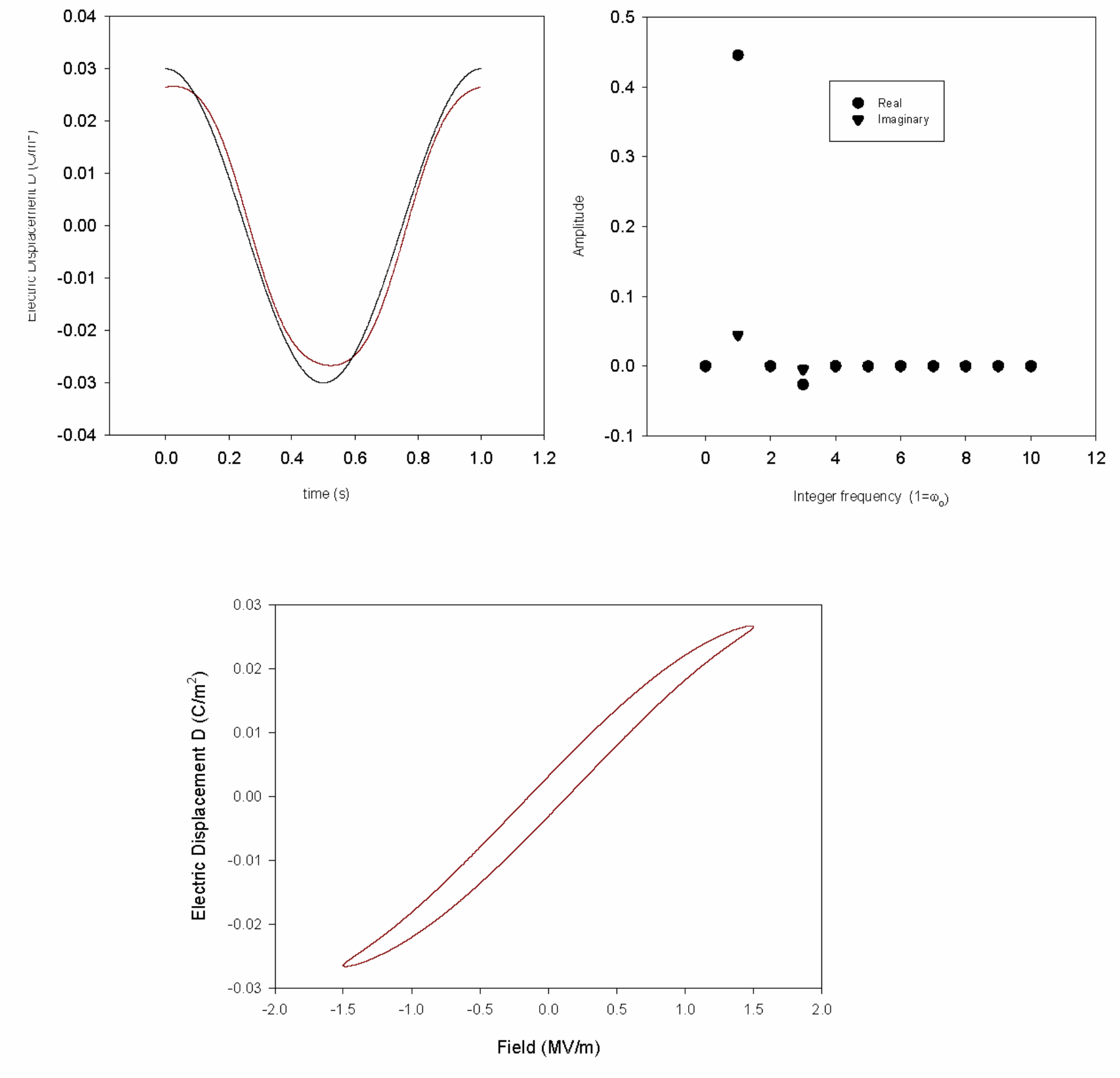

**Fig. 12.** A bipolar non-linear response. Ex. Electric displacement versus field. We have shown the time response along with a cos(ωt) responce and the complex discrete fourier components of D for a harmonic excitation $E(t) = E_0\cos(\omega t)$ with the linear and 3rd order coefficients complex. The bottom curve is the projection on the D(t)-E(t) plane

As can be seen from the measurements shown in Figure 14a, the average constant will be a function of the maximum field applied and will generally increase as the maximum field is increased (to reflect larger irreversible contributions) until saturation is reached. As shown in Figure 13, it is also possible to apply a DC bias field $E_{DC}$ to reach a certain point on the hysteresis curve, then superpose a small AC signal and measure the AC response. The slope of this response would be smaller than the average slope of the hysteresis curve as it would only reflect reversible changes. The slopes determined under bias can be used to characterize materials that do not have a remnant polarization such as the electrostrictive ceramics. When the applied field exceeds the coercive field of the ferroelectric the hysteresis and non-linearities become much more prominent as shown in Figure 14b. This discussion underlines the importance of determining the material constants in keeping with the requirements for any particular application.

To determine the non-linear coefficients it therefore becomes necessary to conduct measurements that determine the polarization and strains directly as a function of applied AC and DC fields and applied stresses. These measurements can be conveniently carried out under near DC conditions. The many different quasi-static measurements that can be carried out can be determined from the sets of constitutive equations (0.3)-(0.6) by putting one of the variables equal to zero and these possibilities are shown in Table 9. Mixed measurements are possible but are difficult in practice given the problem of measuring two variables while adjusting two other variables and these measurements are also difficult to interpret. However it should be noted that there is no reason to expect that the effects are independent. In ferroelectric materials, when the electric field is set to zero and the strain and electric displacement are measured as a function of the stress, the results would be different from the case where a stress and an electric field were simultaneously applied, since both the stress and the electric field can affect the polarization in the material.

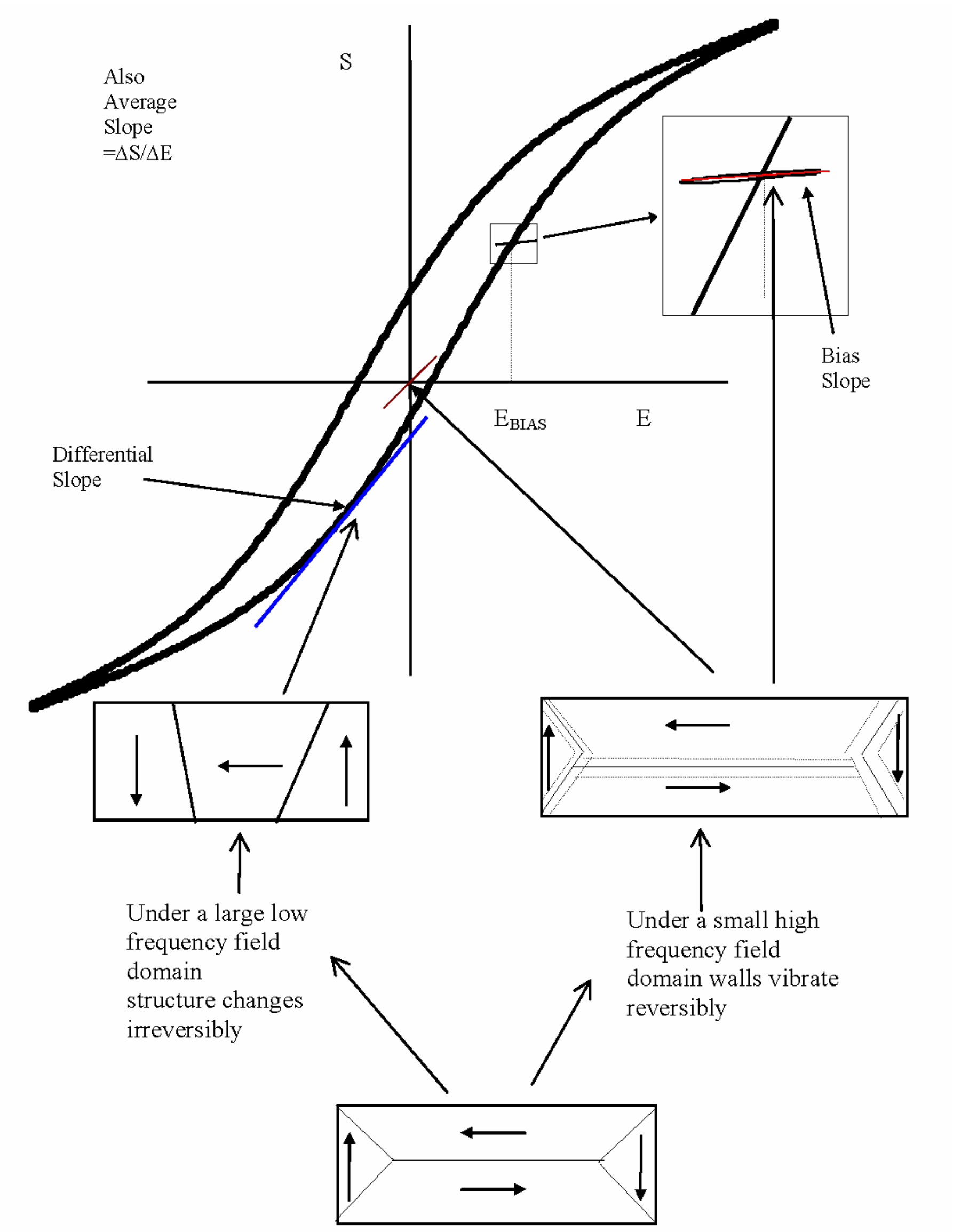

**Fig. 13.** A stylized hysteresis curve to look at the effect of domains on the strain and electric displacement of a piezoelectric ceramic and the various possible measurements of the piezoelectric coefficient.

The most commonly performed quasi-static experiments reflect the conditions described in the two top rows of Table 9. Two experimental conditions are used to simplify the measurement. If the stress is set to zero (i.e. the sample is allowed to expand unhindered), the equations are no longer coupled and the strain and electric displacement may be measured as a function of the electric field. Similarly if the applied electric field is zero (the short circuit condition), the strain and the electric displacement may be measured as a function of stress. These uncoupled relationships are shown in right hand columns of Table 9.

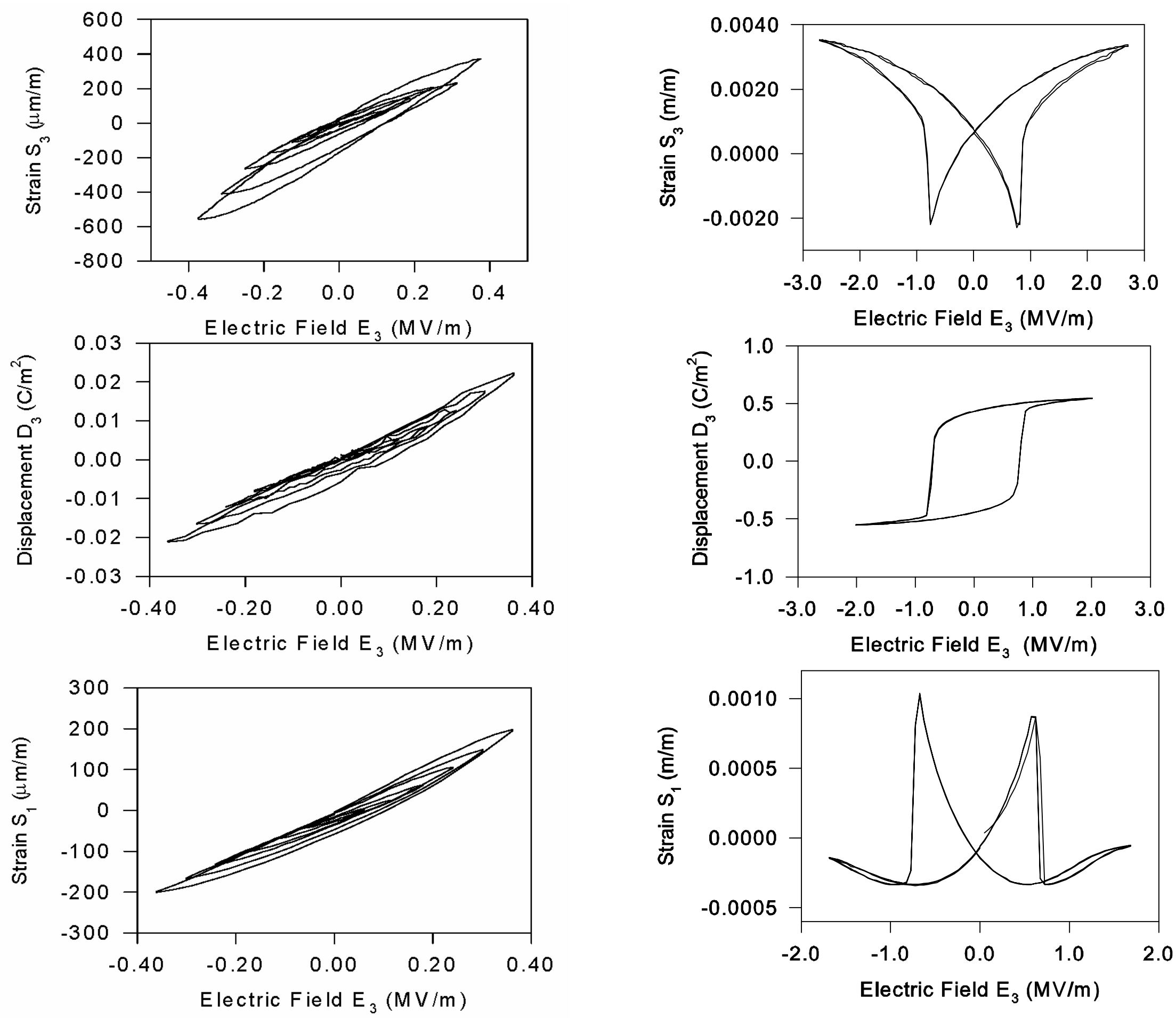

**Fig. 14** The quasi-static field dependence of the strain $S_3$, the electric displacement $D_3$ and the strain $S_1$ in a lead zirconate titanate ceramic as a function of electric field at field levels below the coercive field (left figures) and above the coercive fields (right figures) (Sherrit et al. 1997f).

### *1.4.2 Quasi-static Measurements*

Figure 15 shows a general experimental setup that allows for the simultaneous measurement of the strain $S$ and electric displacement $D$ as a function of an applied electric field $E$. A variety of methods can be used for measuring the strains of piezoelectric materials and these will be discussed below. The electric displacement is usually measured using some variation of the method of Sawyer and Tower (Sawyer and Tower 1930). If the sample area is $A$ and the thickness is $t$ the field experienced by the sample is $E = (V - V_c)/t$ and the electric displacement is $D = CV_c/A$ where $C$ is a standard low loss capacitance in series with the sample and $V_c$ is the voltage measured across this capacitance. The strain $S$ may be found by measuring an analog output of the strain sensor: $S = kV_{\Delta x}$ where $k$ is a linear constant that relates the strain to the output voltage $V_{\Delta x}$ of the strain sensor. Alternately the controller of the strain sensor may have a digital output that can be fed directly to a computer. The various strain measurement techniques are discussed in the next section.

**Table 9.** The quasi-static measurements that can be made on a piezoelectric material with one of the independent variables set to zero.

| Linear Equations | Boundary Conditions | Simultaneous Equations | |
|---|---|---|---|
| $S_p = s^D_{pq}T_q + g_{pm}D_m$<br>$E_m = \beta^T_{mn}D_n - g_{pm}T_p$ | T =0 (unclamped)<br>Apply E – measure S and D | S = dE | D =$\varepsilon^T$E |
| $S_p = s^E_{pq}T_q + d_{pm}E_m$<br>$D_m = \varepsilon^T_{mn}E_n + d_{pm}T_p$ | E = 0 (short circuit)<br>Apply T – Measure S and D | S = $s^E$T | D = dT |
| $T_p = c^E_{pq}S_q - e_{pm}E_m$<br>$D_m = \varepsilon^S_{mn}E_n + e_{pm}S_p$ | T = 0 (unclamped)<br>Apply D (charge electrode) – Measure S and E | S = gD | E = $\beta^T$D |
| $T_p = c^D_{pq}S_q - h_{pm}D_m$<br>$E_m = \beta^S_{mn}D_n - h_{pm}S_p$ , | D = 0 (open circuit)<br>Apply T – Measure S and E | S = $s^D$T | E = -gT |

Figure 16 shows a general setup for measuring the strain *S* and the electric displacement *D* as a function of a transverse stress applied to the specimen. A sheet of sample material is held between grippers. A DC force is used to take up slack in the sample and an AC force is applied. The charge generated in the sample is collected on a large capacitor connected in parallel with the sample. The strain is measured by monitoring the displacement of the free gripper. If a force *F* is applied to a sample of thickness *t*, length *l*, width *w* and electrode length $l_e$, then the stress is $T = F/wt$, the electric displacement is $D = CV_c/wl_e$ and the strain is $S = \Delta x/l$. The frequency of the AC forces that can be applied has a limit that is determined by the mass of the grippers.

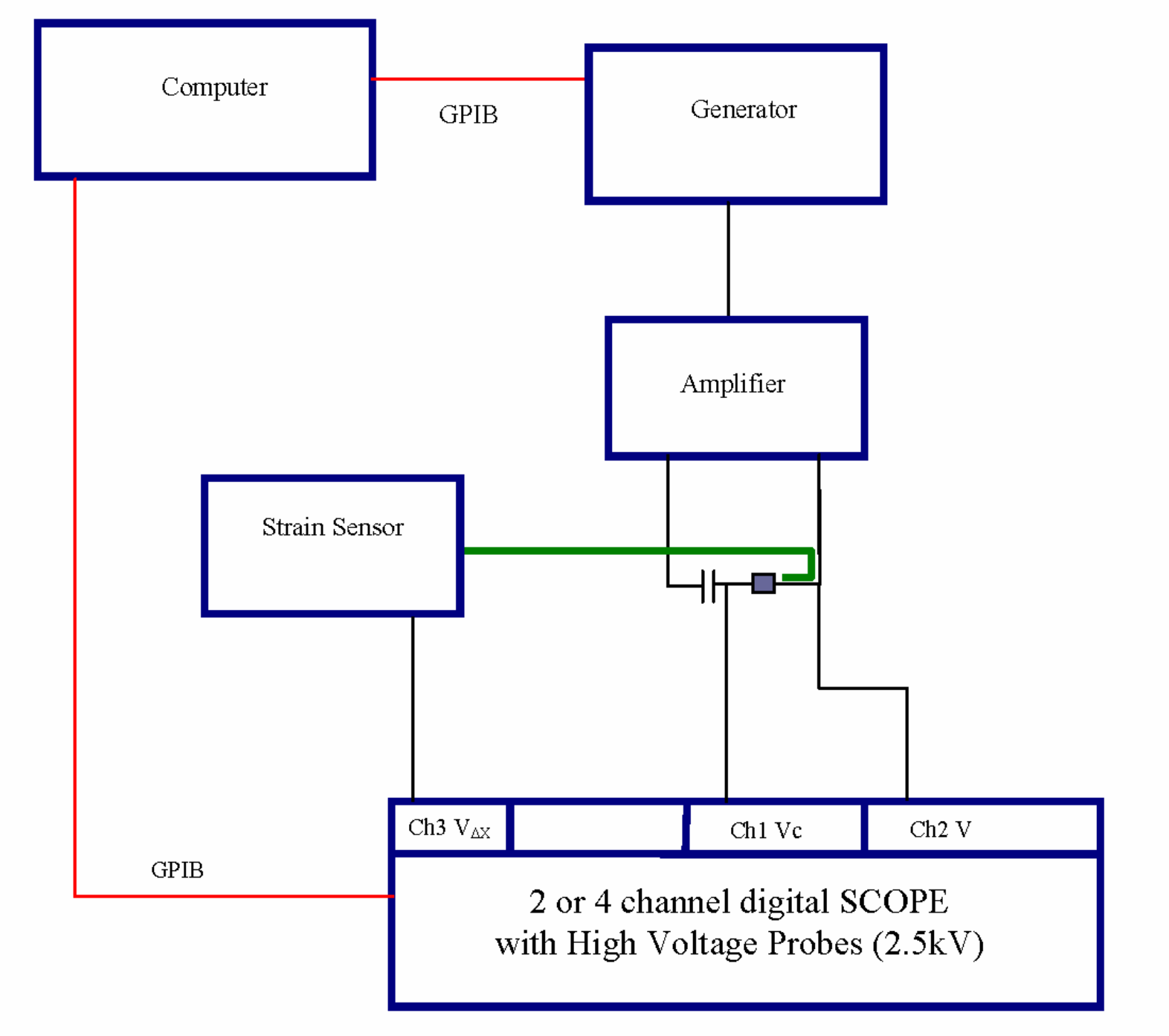

**Fig.15 .** Schematic of a system to measure electric displacement as a function of the electric field (*D* vs. *E*) (Sawyer Tower Circuit) and simultaneously the strain as a function of the electric field (*S* vs. *E*) as discussed by Sherrit (Sherrit et al. 2001a, 2001b).

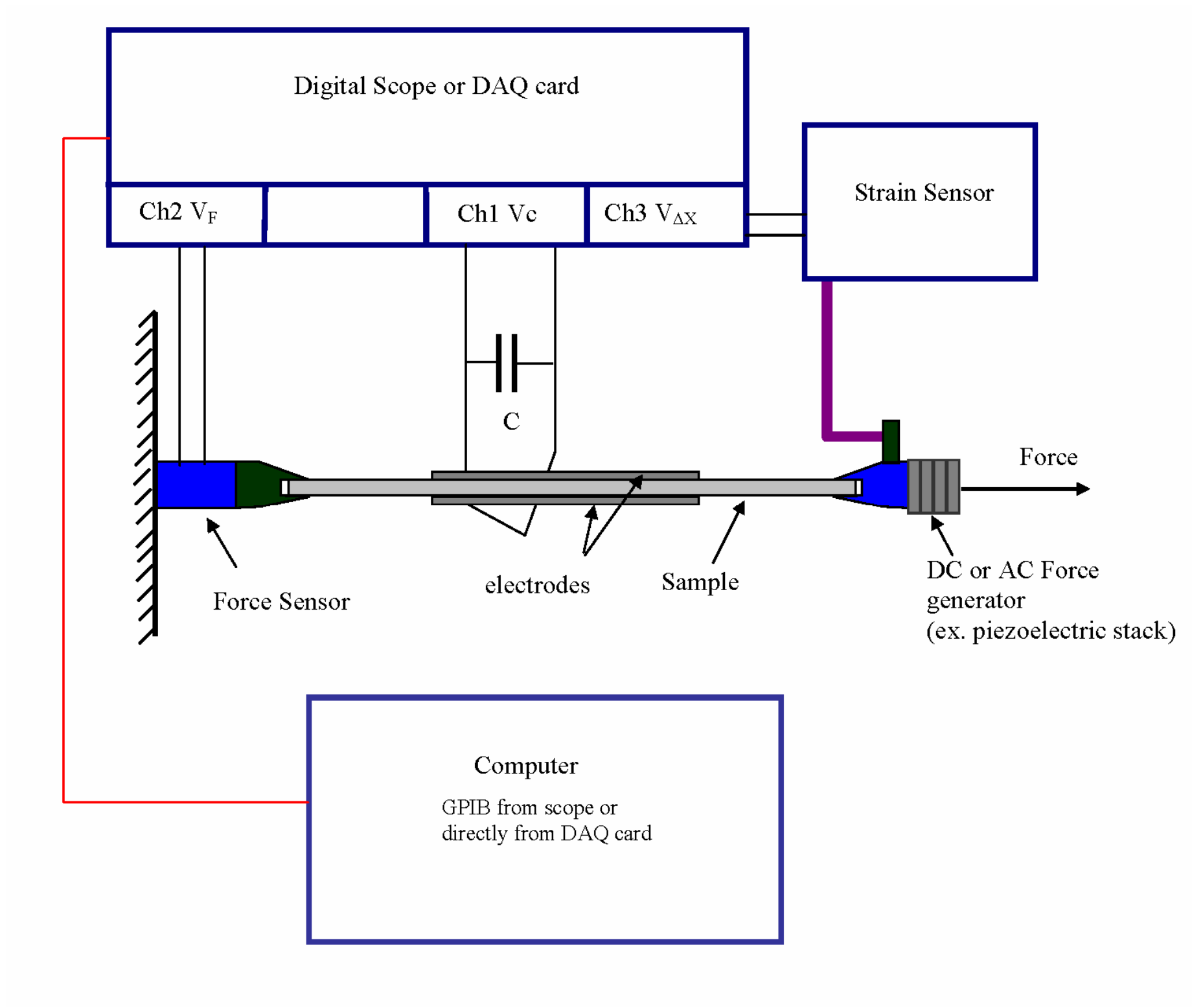

**Fig. 16.** A schematic of an apparatus (Sherrit et al. 2001b) to measure the strain *S* and electric displacement *D* as a function of the transverse stress *T*. Similar arrangements can be used to measure compressive longitudinal stress dependencies as discussed by Sherrit (Sherrit et al. 2001b).

### 1.4.3 Strain measurements

The direct measurement of piezoelectric strain usually involves the measurement of small displacements. A variety of techniques are available to measure the small displacements associated with piezoelectric strain. Optical interferometers have been used by a number of authors (Bruins et al. 1975), (Uchino and Cross 1980), (Pan and Cross 1989) to measure strains at and below the nanometer range. This technique involves monitoring the interference pattern produced by a laser beam and calculating the net displacement from shifts in intensity of the interference pattern. It has a number of problems including the affixing or evaporating of mirrors on to samples, which can be tedious for a large number of samples and may also affect the accuracy of the measurement. The strain range that can be measured is limited by the wavelength of the laser light unless fringe counting is incorporated into the measurement control. These interferometric systems are generally complex and require considerable effort to maintain accuracy. Another technique that has been used is the capacitance dilatometer discussed by (Uchino and Cross 1980) and (Vieira 1986). This technique monitors the capacitance of a parallel plate capacitor and relates the change in capacitance to a dimensional change in the thickness of the capacitor. Problems with this type of measurement include the difficulty in determining parallelism and the initial thickness of the capacitor, thermal drift, vibration suppression and edge effects. A simple and relatively inexpensive way to magnify and measure small piezoelectric strains caused by slowly varying DC fields is to use a lever system. The use of an optical lever system outlined by (Wiederick et al. 1996) is illustrated in Figure 17. A helium-neon laser beam is reflected off the front silvered surface of a 90 degree prism mirror on to a horizontal mirror and back to the rear surface of the prism where it is reflected on to a linear CCD array (or a lateral effect detector). The prism rests on an off centre knife-edge and is balanced by a light metal pin that rests on the sample and acts as a mechanical connection between the prism and the sample. The application of an electric field to the sample causes the sample to deform by $\Delta x$, the mirror is tilted by an angle $\phi = \Delta x/d$ where $d$ is the separation between the supports of the prism. This tilt causes the beam position on the CCD linear array to be shifted by $\Delta s = 4\phi(L_1 + L_2)$ where $L_1$ is the distance from the prism to the CCD and $L_2$ is the distance from the horizontal mirror to the prism. The beam position in the CCD can be monitored with the help of an analog to digital converter card and the displacement of the sample at the point of the support pin is then given by:

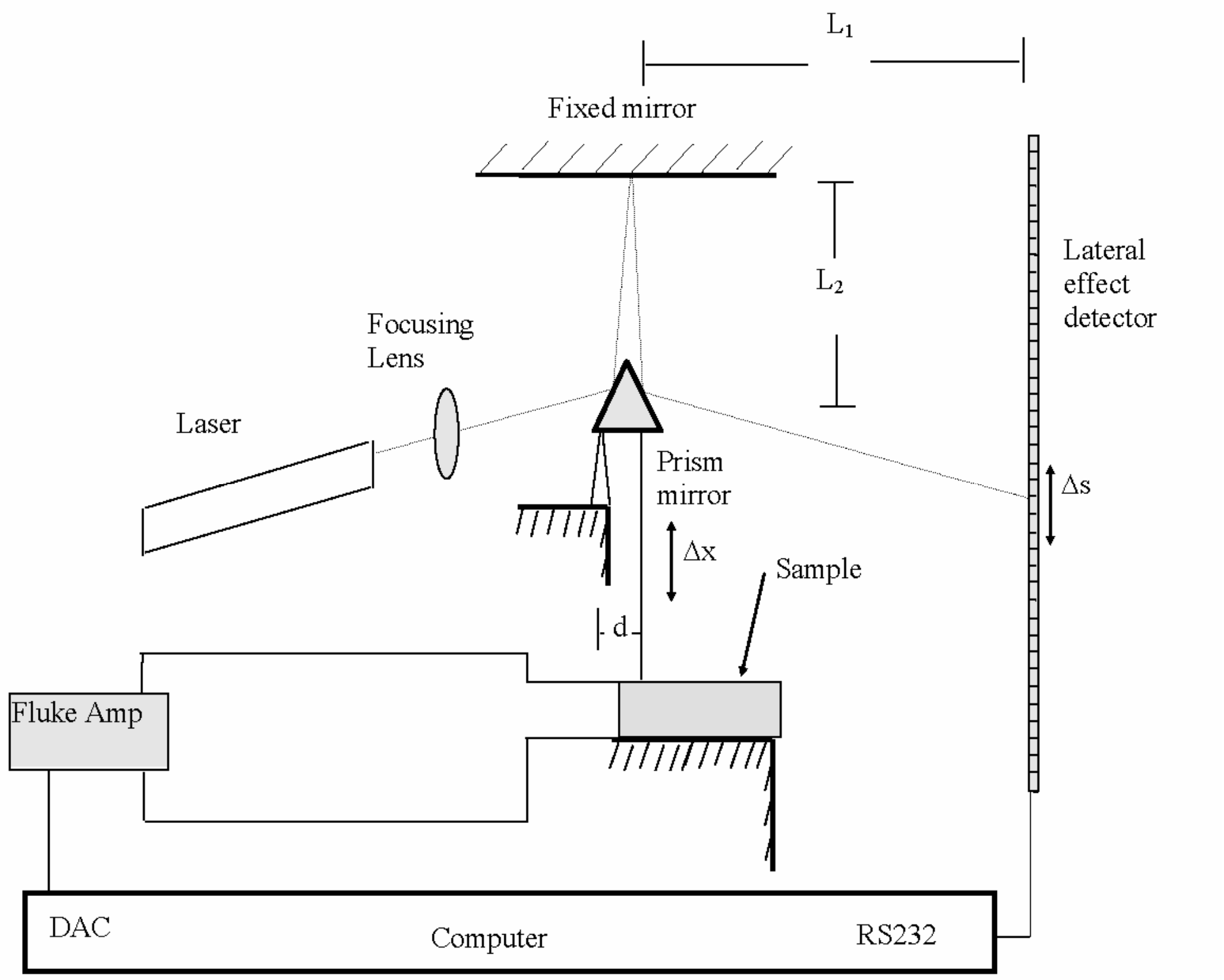

**Fig. 17.** Schematic drawing of the experimental setup for an optical lever measurement of piezoelectric strain (Wiederick et al. 1996). The laser beam's path is shown as a dotted line.

$$\Delta x = \frac{d\Delta s}{4(L_1 + L_2)} \tag{0.106}$$

This experiment is easy to set up and (Wiederick et al. 1996) found that displacements from 2 nm to 5 mm could be made with very little adjustment to the set up. The experimental data shown in Figure 14a and Figure 14b were collected using the optical lever described above. By placing the support pin on different points on the specimen, it is possible to make measurements of the displacements at these different points separately to determine variations in the strain over a particular dimension or in the case of macroscopic composites it is easy to determine the strain in each of the phases.

While the optical lever described above is useful for measuring low frequency strains, higher frequency strains may be measured using commercial contact type displacement transducers such as the linear variable differential transformer (LVDT) or its near relative the differential variable reluctance transducer (DVRT). These are basically electrical transformers with a solid magnetic core that is free to move axially inside the coil's hollow bore. The primary winding is energized by an alternating current and the electrical output signal is the differential voltage between two secondary windings, which varies with the axial position of the core within the coil. This AC output signal is usually converted to a DC signal that is proportional to the movement of the core. These transducers come in different sizes and are made by several manufacturers. They may be used to measure displacements from 50 nm to 50 cm and can nominally measure strains at frequencies up to a few kHz.

Although the pin used in the optical lever and the cores used in LVDTs and DVRTs are usually very light, they may impose a small load on the material and disturb its response. Optical methods may be used for making non-contact measurements of strain and they usually also allow for measurements at higher frequencies. Simple photonic sensors use fiber optic measurement systems and can be used to measure displacements from 1 nm to several millimeters at frequencies ranging from DC to over a 100 kHz. Commercial photonic sensors are available from several manufacturers.

As mentioned above, optical interferometry has been used for non-contact strain measurements. Both homodyne techniques presented by (Uchino et al. 1982) that use fringe counting and heterodyne techniques presented by (Sizgoric and Gundjian 1969) and (Kwaaitaal 1974) that compare a measured unknown frequency to a reference frequency have been used. The heterodyne technique has the advantages of phase detection, wide bandwidth, high stability and easy optical alignment. The ZMI 2000 laser Doppler interferometer manufactured by Zygo Corporation uses a heterodyne technique and can be used to measure extensional, transverse and shear strains as a function of applied electric fields on a piezoelectric material (Masys et al. 2003). A sine wave voltage generated by a function generator is amplified by a power supply/amplifier and then applied to the ceramic sample. The polarized beam from a helium neon laser has two frequency components $f_1$ and $f_2$ that are orthogonally polarized with a frequency difference of 20 MHz. The beam is split into two beams by a beamsplitter (BS). One beam goes through interferometer 1 and only the component $f_1$ is incident on the top surface of the sample and is used to measure the longitudinal displacement. The second beam goes through interferometer 2 and the beam component $f_1$ is incident on the side surface of the sample and is used to measure transverse displacement. A frequency shift $\Delta f_1$ in the beam reflected by a surface moving at velocity $v$ is given by the equation:

$$\Delta f_1 = \frac{2v}{\lambda} \tag{0.107}$$

where λ is the laser wavelength, 633 nm. The beam that is reflected from the sample with a frequency of $(f_1+\Delta f_1)$ is recombined with the frequency $f_2$ in the interferometers and the output beam $\{f_2 - (f_1+\Delta f_1)\}$ is sent to a phase detector along with a reference beam $(f_2 - f_1)$ through optical fibers. The decoded frequency shift $\Delta f_1$ is converted to a voltage that is proportional to the velocity of the sample surface.

The displacement and the strain of the sample can then be calculated. The entire system may be computer controlled for automation of the measurement. A small optically flat mirror is attached to the sample surface to provide a good reflecting surface. The sample is attached to a sample stage with a small amount of epoxy and the sample is aligned to achieve maximum signal strength of the reflected beam. In order to achieve high electric fields it is usual to use fairly thin piezoelectric samples and care should be taken to prevent discharges when high AC or DC voltages are applied across the thin specimens. Care should also be taken to ensure that thin flat specimens do not bend as a result of the applied field. The two beams of the system allow for the possible determination of all the three piezoelectric coefficients $d_{33}$, $d_{31}$ and $d_{15}$ by adjusting the applied field and reflecting surfaces. Other interferometers such as the Doppler interferometer are available that promise much higher frequency responses.

### *1.4.4 Biased Resonance measurements*

Another useful approach especially with electrostrictive materials which have a field activated piezoelectric coefficient is biased field resonance measurements. Under bias the electrostrictive material behaves as a piezoelectric with the magnitude of the piezoelectric coefficient initially being linear with field. In order to describe this behavior we need to understand the macroscopic behavior of these materials. The macroscopic modeling of these materials has received much attention. (Hom et al. 1994) use a set of equations relating the stress $T_{ij}$ and Polarization $P_i$ to the strain $S_{ij}$ and electric field $E_i$. In their model they use the *tanh* function saturation model proposed by (Zhang and Rogers 1993) to describe the polarization at high fields. (Piquette and Forsythe 1997) have put forth an alternative model based on saturation in P that goes as E divided by the square root of a quadratic in E. An alternate derivation in $T_{ij}$ and $E_i$ based on the Gibbs' function has recently been published by (Kloos 1995a, 1995b). Let us examine the behavior of the PMN-PT-La (0.9/0.1/1%) relaxor ferroelectric material with respect to the electrostriction model developed by (Mason 1958). Mason's derivation from the elastic Gibbs' potential (strain S and electric field E as a function of the electric displacement D and stress T) is for a reversible material with zero hysteresis.

The equations for the electric field $E_i$ and strain $S_{ij}$ proposed by Mason are:

$$E_i = -2Q_{klij}T_{kl}D_j + \left[\beta^T_{ij} + R_{ijmnkl}T_{mn}T_{kl}\right]D_j + \Sigma_{ijkl}D_jD_kD_l + \Sigma_{ijklmn}D_jD_kD_lD_mD_n$$
$$S_{ij} = \left[s^D_{ijkl} + R_{ijmnkl}D_mD_n\right]T_{kl} + Q_{ijmn}D_mD_n \quad (0.108)$$

where $Q_{klij}$ is the electrostriction coefficient relating the strain to the electric displacement. $\beta^T_{ij}$ is the inverse permittivity, $R_{ijmnkl}$ is the morphic correction term for the dependence of the open circuit elastic constant $s^D_{ijkl}$ on the electric displacement and $\Sigma_{ijkl}$ and $\Sigma_{ijklmn}$ are higher order inverse permittivity terms (anisotropic energy terms) which account for the increase in field at higher electric displacement (which are normally thought of in terms of the inverse relationship as the saturation in $D_i$ as a function of $E_i$). It should be noted that, due to the higher order nature of these equations, transformation equations to other data representations (E or T, E or S, and D or S as the intrinsic variable) like those found in linear piezoelectric theory are only directly solvable in some limited situations.

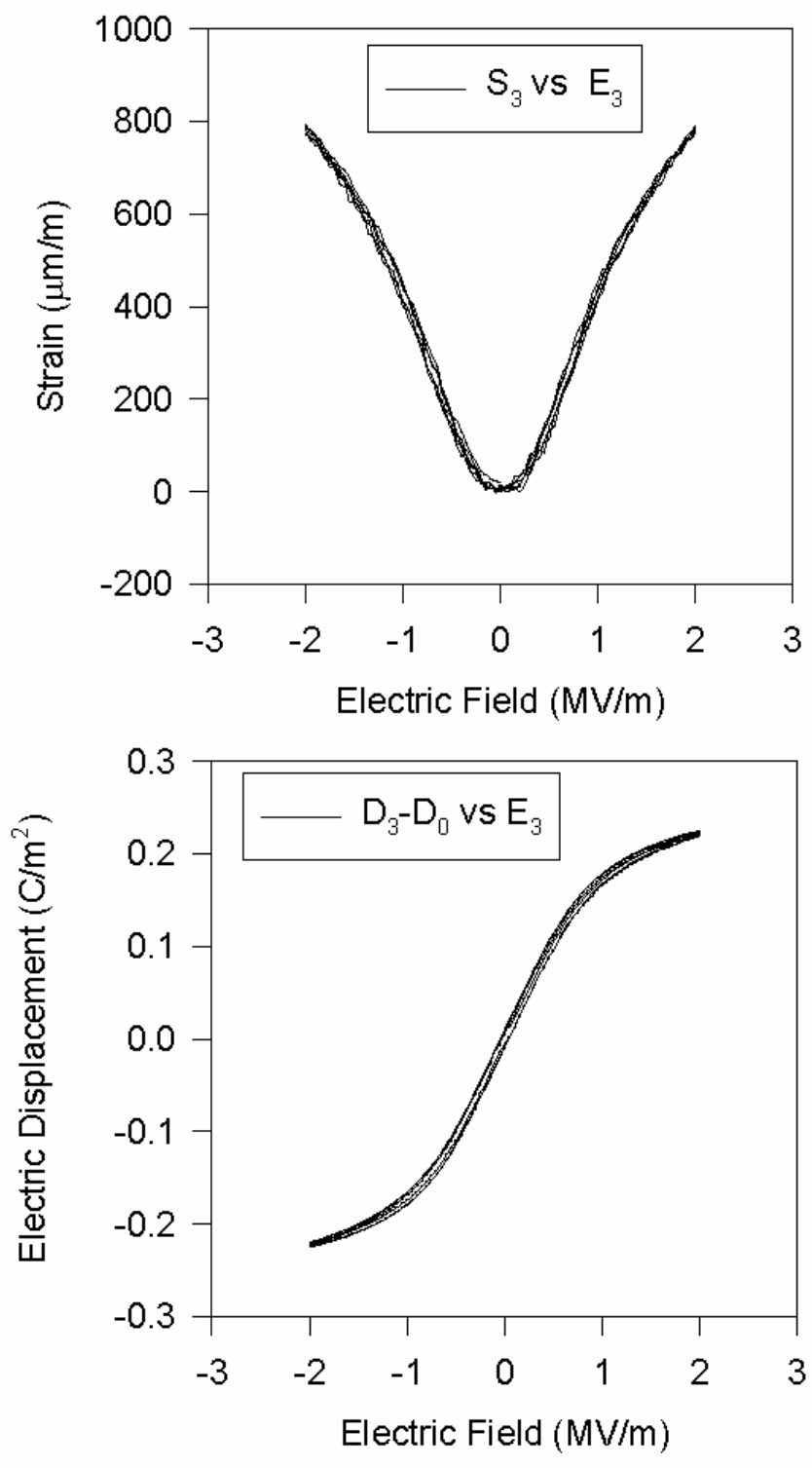

**Fig. 18.** The strain $S_3$ and electric displacement $D_3$ for a PMN/PT/La (0.9/0.1/1%) sample under zero load. Strain is an even function of E while the electric displacement corrected for $D_0$ is an odd function in field E.

As well it is apparent from the coupled equations shown in equation (0.108) that when the data is transformed to S and D versus T and E that saturation in strain as a function of the electric field will result from the higher order permittivity terms. Consider a stress free system under the application of an electric displacement in the 3 direction in reduced notation Mason's equations reduce to.

$$E_3 = \beta^T_{33}D_3 + \Sigma_{3333}D_3^3 + \Sigma_{333333}D_3^5$$
$$S_3 = Q_{333}D_3^2 \quad (0.109)$$

The inverse of the first equation shown in equation (0.109) is also an odd polynomial and can be written as

$$D_3 = \varepsilon^T_{33}E_3 + \kappa_{3333}E_3^3 + \kappa_{333333}E_3^5 \quad (0.110)$$

where the relationship between the coefficients in the D vs E and E vs D can be determined independently by fitting the data in each representation. Since a power of E or D can be factored out of each of these equations the equation is only a fourth order and a direct transform from one equation set to the other may be possible however for complete saturation higher order odd terms are required. The equation for the strain as a function of the electric field is found by substitution of equation (0.110) into the second equation of equation (0.109). The equation for the strain is then

$$S_3 = Q_{333}\left[\varepsilon_{33}^T E_3 + \kappa_{3333}E_3^3 + \kappa_{333333}E_3^5\right]^2$$
$$= Q_{333}\left(\varepsilon_{33}^T\right)^2 E_3^2 + 2Q_{333}\kappa_{3333}\varepsilon_{33}^T E_3^4 + Q_{333}\left(\kappa_{3333}^2 + 2\kappa_{333333}\varepsilon_{33}^T\right)E_3^6 + ....... \quad (0.111)$$

where we see that the saturation in strain as a function of the electric field is essentially due to the higher order permittivity terms that result from the saturation in D with increasing E ($\kappa_{3333}$ term is negative). From this relationship it is also easy to see why the effective piezoelectric coefficient under a bias field is found to reach a maximum and then decrease.

**Table 10.** Typical coefficients of the odd and even polynomials fit to the PMN-PT-La (.9/.1/1%) data shown in Figure 21

| Variable | Polynomial (E in V/m) |
|---|---|
| $S_3$ (m/m) | $= 6.0\text{x}10^{-16}E^2 - 1.9\text{x}10^{-28}E^4 + 2.3\text{x}10^{-41}E^6$ |
| $D_3$ (C/m$^2$) | $= 2.2\text{x}10^{-7}E - 5.0\text{x}10^{-20}E^3 + 6.0\text{x}10^{-33}E^5$ |

Typical strain-displacement data for the PMN/PT/La (0.9/0.1/1%) material is shown in Figure 18. The coefficients of the polynomial fit to the data using equations (0.110) and (0.111) are shown in Table 10. The 4th order term in field in the strain equation is negative and is responsible along with the higher order even coefficients for the saturation in strain as a function of the electric field. The effective piezoelectric $d_{33}$ constant from equation (0.111) is then calculated to be

$$d_{33} = \frac{dS_3}{dE_3} = 2Q_{333}\left(\varepsilon_{33}^T\right)^2 \bar{E}_3 + 8Q_{333}\kappa_{3333}\varepsilon_{33}^T\bar{E}_3^3 + 6Q_{333}\left(\kappa_{3333}^2 + 2\kappa_{333333}\varepsilon_{33}^T\right)\bar{E}_3^5 + ....... \quad (0.112)$$

where we have used the bar above the electric field to signify DC bias. The maximum value of the effective $d_{33}$ coefficient can now be found by setting the derivative of equation (0.112) with respect to field to zero and solving for E. Taking only the first two terms in equation (0.112) we find the field $E_{max}$ at which the effective $d_{33}$ is a maximum is $E_{\max} = (\varepsilon_{33}^T/12\kappa_{3333})^{1/2}$. Using the values from Table 10 we find the field at which the piezoelectric constant is a maximum is 0.61 MV/m. When all terms are included the maximum in the piezoelectric coefficient is found to occur at 0.70 MV/m which is in reasonable agreement with experimental data. It is interesting to note that the piezoelectric voltage coefficient $g_{33}$ can be shown not to reach a maximum but rather saturate in field with the same field dependence as the electric displacement ($g_{33} = 2Q_{333}D_3$). A similar relationship is found to hold for the $e_{33}$ (reaches maximum) and $h_{33}$ (saturates) coefficients.

The PMN-PT-La (0.9/0.1/1%) material has been found to have a small remanent electric displacement $D_0$ which produces a small but measurable hysteresis and a linear component to the strain electric displacement ( $S = QD^2$ ) curves. This linear component ( $S = -g|D|$ ) is found to switch sign as the displacement passes near the origin (Sherrit et al. 1998b) The total strain which is the superposition of the quadratic electrostrictive component and the switching linear component is an even function of the electric displacement. This means that the total strain can be modeled as the sum of higher order even polynomials in D, however the value of the total quadratic component is no longer the true electrostriction since the resultant quadratic component is the sum of the true electrostriction and the quadratic component of the expansion of the switching linear component (Sherrit et al. 1998b).

As can be seen from equation (0.112) an induced piezoelectric behavior under a DC bias is found which has significant implications for the generation of ultrasound. The derivation of the field dependence of the small signal parameters of these materials can be shown formally by expanding the appropriate thermodynamic potential to arbitrary field, stress, electric displacement or strain. To determine the field dependence of the small signal parameters for a given resonance mode the derivative of the coupled field equations is taken with respect to the intrinsic variables and the terms are gathered in common differentials. Then the boundary conditions are applied and the resonance equations can be calculated in terms of the material coefficients which are now functions of the DC field. For the length extensional resonator, for example, we find.

$$s_{33}^E\left(\bar{E}_3\right) = s_{33}^E + 2\varsigma_{333}\bar{E}_3 + \tau_{3333}\bar{E}_3^2 \quad (0.113)$$

$$d_{33}\left(\bar{E}_3\right) = d_{33} + 2\gamma_{333}\bar{E}_3 + 3\psi_{3333}\bar{E}_3^2 + 4\chi_{33333}\bar{E}_3^3 \quad (0.114)$$

$$\varepsilon_{33}^{T}\left(\bar{E}_3\right)=\varepsilon_{33}^{T}+2\kappa_{333}\bar{E}_3+3\kappa_{3333}\bar{E}_3^2+4\kappa_{33333}\bar{E}_3^3+5\kappa_{333333}\bar{E}_3^4 \tag{0.115}$$

The electromechanical coupling constant for the length extensional resonator under DC bias is then

$$k_{33}^{2}\left(\bar{E}_3\right)=\left(\frac{d_{33}^{2}\left(\bar{E}_3\right)}{s_{33}^{E}\left(\bar{E}_3\right)\varepsilon_{33}^{T}\left(\bar{E}_3\right)}\right) \tag{0.116}$$

It should be noted that we could have equally used the model of (Hom et al. 1994) or (Piquette and Forsythe 1997) for the electric displacement and strain in the longitudinal mode. An example of elastic compliance and the piezoelectric coefficient determined from a PMN-PT-La (.9/.1/1%) length extensional resonator under bias is shown in figures on the left in Figure 19. The various electromechanical coupling constants as a function of the bias field for this material are shown in the graph on the right in Figure 19. This approach can also be used on piezoelectrics to determine the change in the small signal properties under electric bias fields.

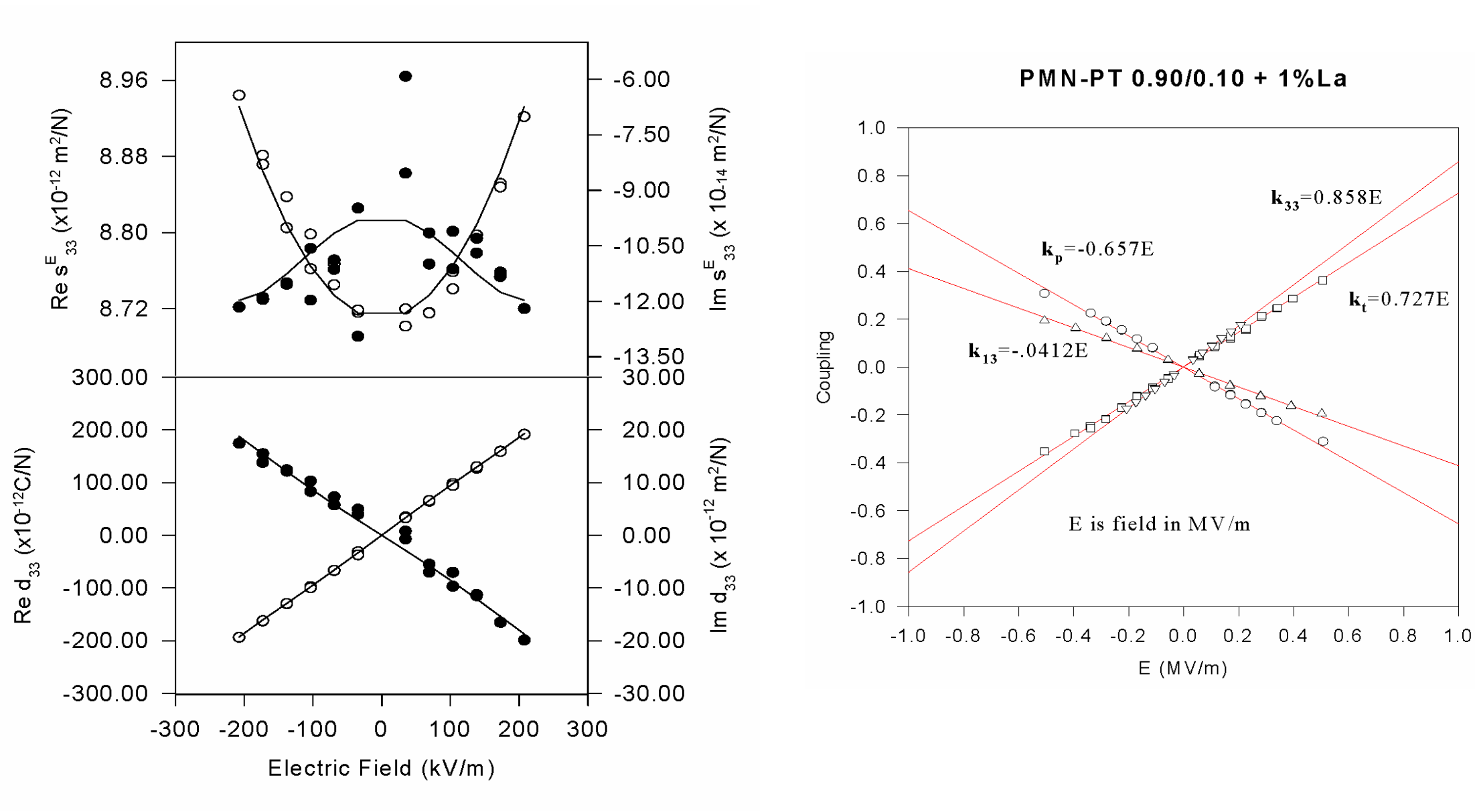

**Fig. 19.** The small signal length extensional material properties for a PMN/PT/La (0.9/0.1/1%) sample under zero load are shown in the two graphs on the left. The real and imaginary components of the elastic compliance and effective piezoelectric constant of the LE (l=7.9mm, w1=1.47mm, w2=1.45mm, density =7680 kg/m3) mode for the PMN/PT/La 0.9/0.1/1% material as a function of the DC bias field. The open circles represent the real coefficient and the filled circles show the imaginary coefficient. The left figure shows the electromechanical coupling of the various modes for the PMN/PT/La 0.9/0.1/1% material below the onset of saturation which begins between 0.4 and 0.5 MV/m. The linear component of the coupling is shown for each mode.

### *1.4.5 Resonance measurements under High Drive*

High power density ultrasonic transducers are used in a variety of devices for drilling (Sherrit et al. 2001c), cleaning (Gallego-Juarez 1994), welding (Tsujino 1999) , filters (Larson et al. 2000), and novel transformers (Bove et al. 2000). Driving piezoelectrics at resonance under large fields introduces a host of problems for measuring and interpreting the data. In general driving piezoelectrics at high power levels introduces non-linear and thermal effects that need to be understood in order to accurately measure and control the transducer. In the small signal regime piezoelectrics are typically linear, isothermal, and adiabatic. When the same materials are driven at higher drive levels the thermal conditions can change considerably and the sample under these conditions is no longer isothermal or adiabatic since the large voltages produces self heating (Sherrit et al. 2001a) that can in some circumstances be catastrophic (leads melting, thermal runaway, sample cracking etc.). By definition the standard impedance measurement of the sample is a linear property. It is the ratio of the phasor of the drive voltage at frequency ω divided by the phasor of the current at the drive frequency ω. This defi-nition of impedance assumes no harmonics are generated and calculating the impedance using this definition in essence filters out the harmonics. There are a variety of ways the impedance can be measured and care must be taken to understand the actual technique (sine correlation – Solatron, four-terminal-pair (4TP) auto-balancing bridge –Agilent, etc.) and how the instrument deals with harmonics. Typically harmonics are not generally a problem in commercial impedance meters since they usually limit the drive voltage to order of 1 volt and for most piezoelectric sample geometries the fields produced on the sample are still in the linear regime. An example of a test set-up (Sherrit et al. 2001a) used to investigate non-linearities and thermal issues is shown in Figure 20. All the data can be acquired as a function of time so the frequency components and harmonics can be evaluated. The temperature as a function of the frequency and the admittance for a hard PZT sample (Channel 5800) is shown in Figure 21.

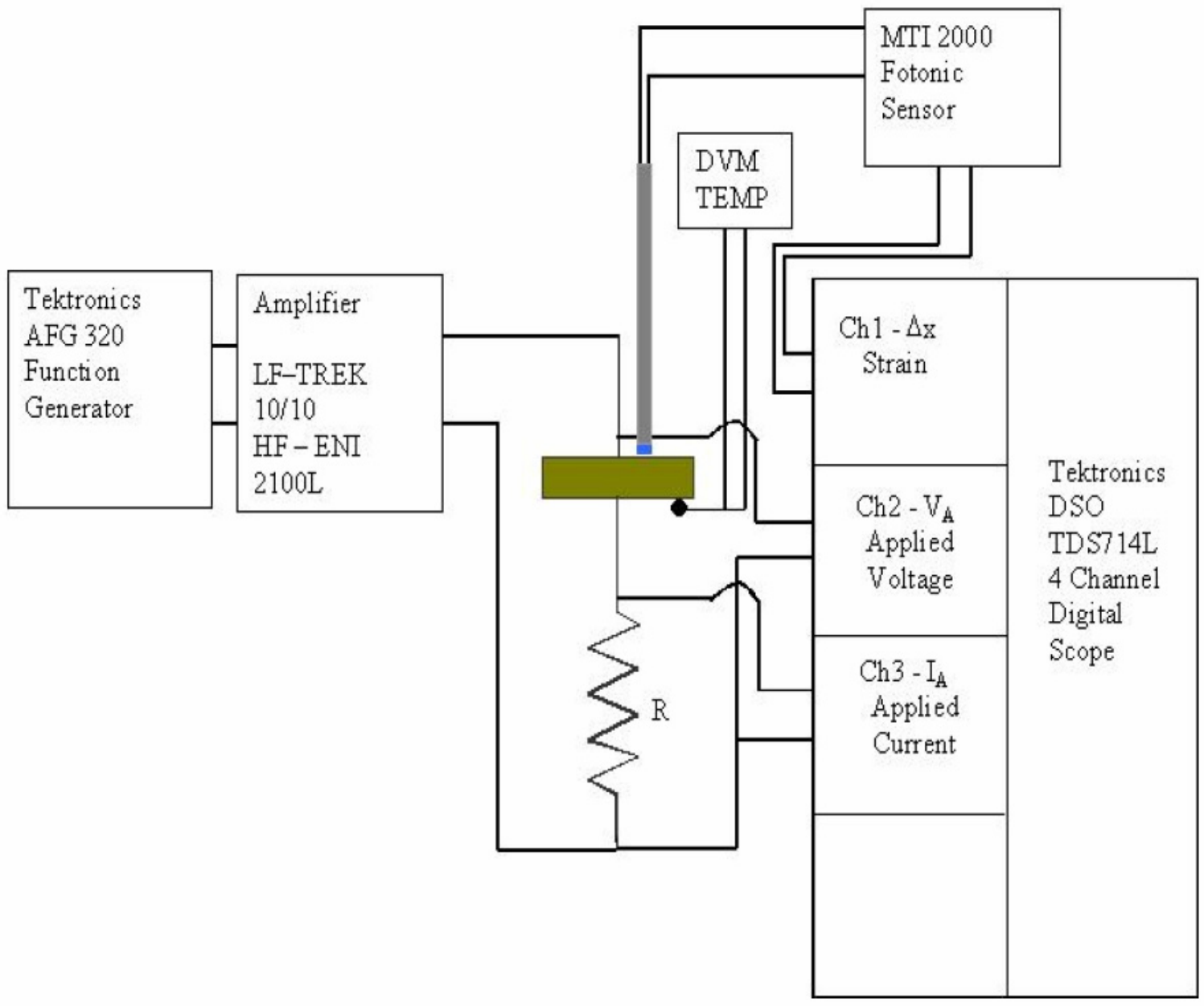

**Fig. 20.** Experimental system to measure the electric field E, displacement D, strain S, temperature T and impedance Z as a function of frequency (Sherrit et al. 2001a). Temperature is measured with a thermocouple.

The data shows a difference in the admittance and temperature profiles depending on whether the frequency is swept up or swept down including a shift in the resonance frequency (frequency hopping). In this case the spectrum was evaluated at each frequency point and the harmonics in the current were ignored. In addition the measurement was continuous hence the good correlation to the temperature frequency curves. In order to separate out the non-linearities from the thermal effects the measurement could be duty cycled to reduce the overall temperature rise, however at high fields and at resonance the heat is still being created and flowing so the adiabatic boundary condition is suspect. Other authors have reported these thermal effects (hysteresis and frequency jumps) (Woolett and Leblanc 1973), (Priya et al. 2001) and attribute the various effects to elastic nonlinearities. It is apparent from the data in Figure 21 that at least part of the hysteresis and frequency jump is due to the lack of isothermal conditions during the measurement. A recent paper by (Blackburn and Cain 2007) suggests the use of a burst method to remove the temperature effects and fit the current and voltage data. Their method demonstrated frequency hopping and tilting of the resonance curve which suggests the changes occurred at a fast enough rate before heat can flow to the different parts of the sample maintaining the adiabatic boundary condition (Nye 2003).

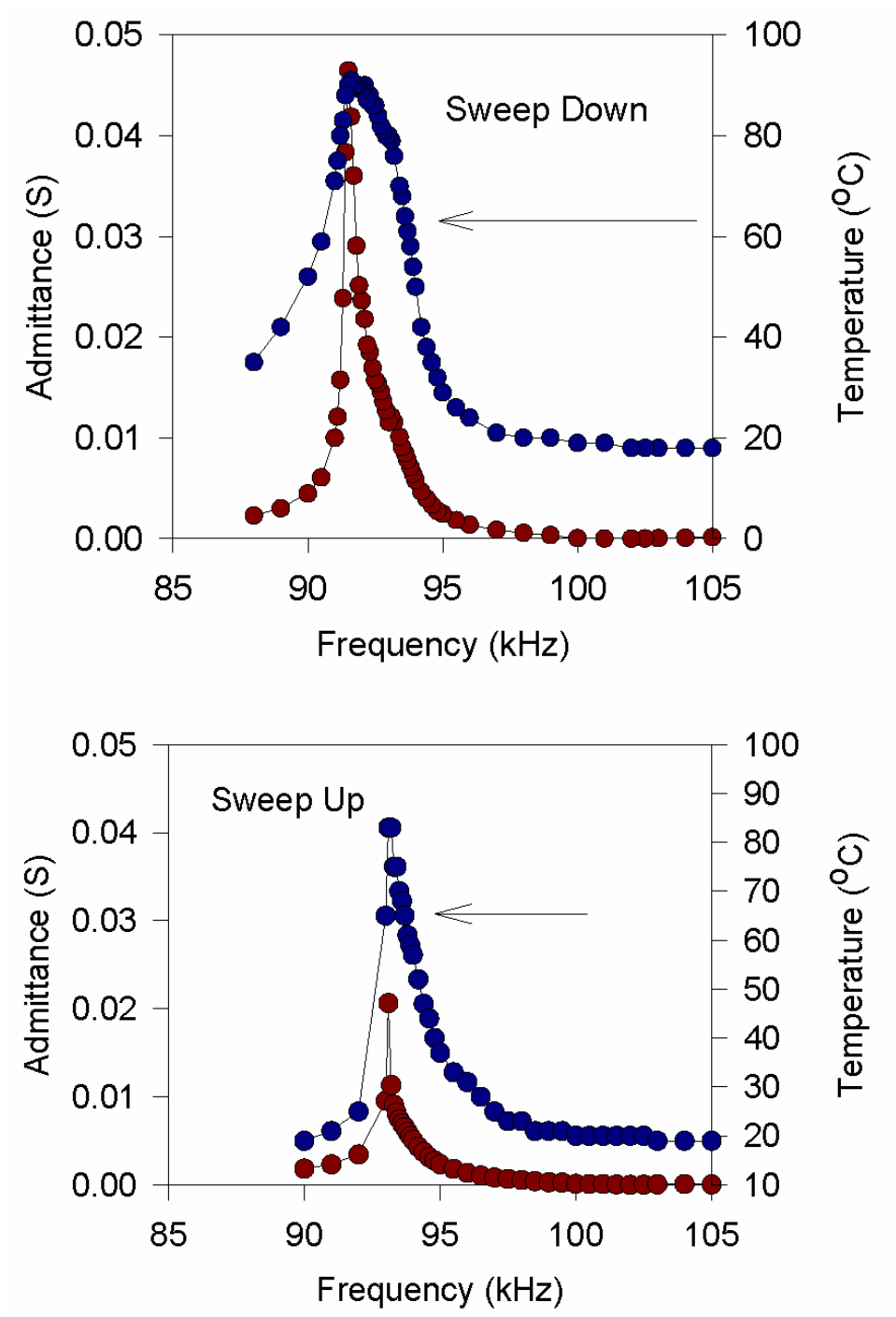

**Fig. 21** Thermal hysteresis and frequency jumping for a ring resonator sample of hard PZT (Channel 5800) around resonance. Frequency shift in the peak of the temperature and admittance are seen to correlate.

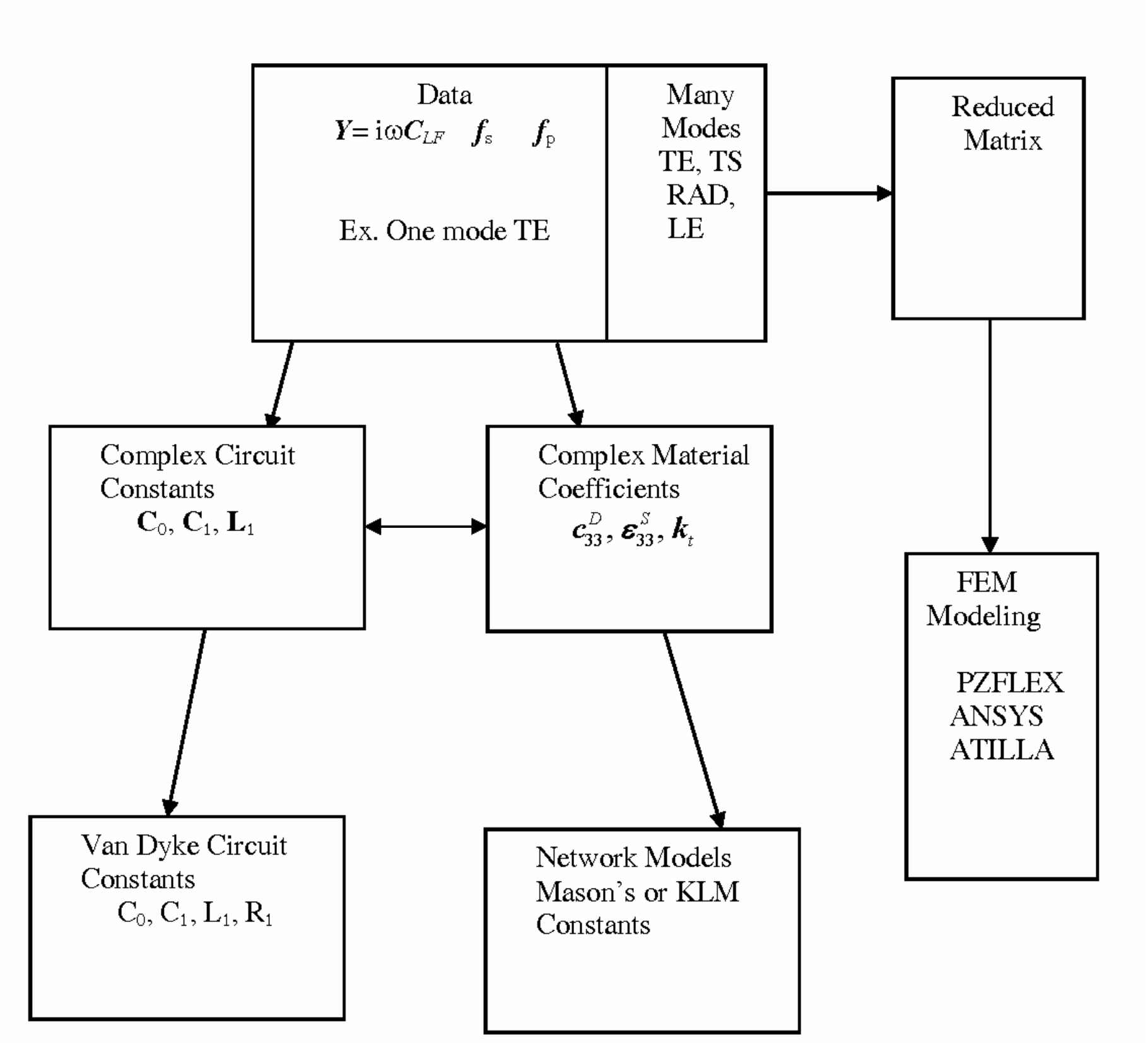

**Fig. 22.** A schematic diagram looking at how the resonance data can be used. It should be noted that the Data representation, Complex Circuit representation and the complex material coefficients are derivable from each other and are exact.

### 1.5 Summary

The schematic diagram shown in Figure 22 outlines the usefulness and the utility of the complex coefficients we have discussed and where they may be used in modeling. From the complex data constants ($\boldsymbol{Y}$= i$\omega \boldsymbol{C}_{LF}$, $\boldsymbol{f}_s$, $\boldsymbol{f}_p$) one can calculate the complex material constants $\boldsymbol{c}_{33}^{D}, \boldsymbol{\varepsilon}_{33}^{S}, \boldsymbol{k}_t$ or the complex circuit constants $\mathbf{C}_0$, $\mathbf{C}_1$, $\mathbf{L}_1$. From the complex circuit constants one can collapse the 3 losses of the complex circuit into one loss $R_1$ to produce the 4 Butterworth Van-Dyke circuit constants $C_0$, $C_1$, $L_1$, $R_1$. In addition the parameters of the Mason's or the KLM network models can be determined from the complex material constants. If the transducer or actuator is not of a standard geometry or has features that make it very difficult to model as an isolated mode a finite element analysis could be performed if all the material constants of the reduced matrix were known. If the finite element package does not handle material damping directly one could use the mechanical quality factors to guide them in the choice and amplitude to the mechanical damping models. Similar approaches can be used to account for the dielectric loss (Powell et al. 1997).

In addition to the methods we have discussed to determine how these materials respond to certain stimuli there are also practical matters that relate to the stability of the transducer or actuator. These properties include; aging, creep, degradation and life. In each of these cases we determine some metric, coefficient or figure of merit and monitor its value as a function of the time after poling (aging) or as a function of the time after applying some field or stress (creep) or as a function of the number of cycles (degradation) or finally until the property degrades past a set limit (life). The choice of the coefficient or groups of coefficients to monitor depends on how the transducer will be used.

In closing we would like to emphasize the need to characterize the material properties under the conditions they will be subjected to in the device or transducer since the properties and stability in some field or stress regimes can change significantly from the small signal values.

ACKNOWLEDGMENTS

S.S. would like to thank Dr. Yoseph Bar-Cohen, Dr. Xiaoqi Bao, Dr. Zensheu Chang, Dr. Mircea Badescu and Dr. Harvey Wiederick for useful discussions. The research at the Jet Propulsion Laboratory (JPL), a division of the California Institute of Technology, was carried out under a contract with the National Aeronautics Space Agency (NASA).

B.M. would like to thank his research associates and students, Dr. Harvey Wiederick, Dr. Guomao Yang, Dr. Wei Ren, Mr. Shi-Fang Liu, Major Anthony Masys and Captain Georges Sabat for their contributions to the research. The work at the Royal Military College of Canada was funded mainly by contributions from the US Office of Naval Research and the Canadian Department of National Defence.